\documentclass[fleqn,usenatbib]{mnras}

\usepackage{newtxtext,newtxmath}

\usepackage[T1]{fontenc}

\DeclareRobustCommand{\VAN}[3]{#2}
\let\VANthebibliography\thebibliography
\def\thebibliography{\DeclareRobustCommand{\VAN}[3]{##3}\VANthebibliography}

\usepackage{graphicx}	
\usepackage{amsmath}	
\usepackage{hyperref}

\title[SEISNe progenitors with BPASS]{The progenitors and circumstellar environments of stripped-envelope interacting supernovae from BPASS}

\author[B. Warwick et al.]{
B. Warwick,$^{1}$\thanks{E-mail: ben.warwick@warwick.ac.uk}
J. Lyman,$^{1}$
D. L. Coppejans,$^{1}$
C. M. Byrne,$^{1}$
J. J. Eldridge,$^{2,1}$
M. Pursiainen,$^{1}$
E. R. Stanway$^{1}$
\\
$^{1}$Department of Physics, University of Warwick, Gibbet Hill Road, Coventry, CV4 7AL, UK\\
$^{2}$Department of Physics, University of Auckland, Private Bag 92019, Auckland, New Zealand\\
}

\date{Accepted XXX. Received YYY; in original form ZZZ}

\pubyear{\the\year{}}

\begin{document}
\label{firstpage}
\pagerange{\pageref{firstpage}--\pageref{lastpage}}
\maketitle

\begin{abstract}
Understanding the progenitors of stripped-envelope interacting supernovae (SEISNe) is crucial for probing the final stages of massive star evolution. Despite this, their rarity means the nature of their progenitors remains poorly constrained. We investigate the progenitors of SEISNe, using the Binary Population and Spectral Synthesis stellar evolution models. The 30,153 stellar models that result in hydrogen-poor core-collapse supernovae (SN) includes both binary and single stars, spanning 13 metallicities ($Z=10^{-5}$ to $0.04$). Circumstellar material (CSM) formation during their last 100\,kyr is reconstructed from line-driven and Roche lobe overflow (RLOF) mass loss using three scenarios for the CSM, one looking at a wind only distribution for the CSM, and two involving the formation of a circum-binary disc (CBD). Light curve parameters from each scenario are calculated using an analytical model and fiducial SN explosion parameters. We find only the two CBD scenarios, and consequently no single star models, reproduce the luminosities and rise times of observed SEISNe. The inferred rates are comparable to observations. Expected progenitors were only found in models with ZAMS masses of $14-40\,$M$_{\odot}$ at $\geq Z_\odot$ and $30-40\,$M$_{\odot}$ at $< Z_\odot$. Modelling the radio emissions shows that early ($\leq8\,$days) and high frequency ($\geq70\,$GHz) observations are required to constrain the nature of CBDs in these systems. These results indicate that, without phenomena such as eruptive mass loss, SEISNe require massive stars in binary systems in order to produce sufficient masses of CSM and confine them close to the progenitor. 
\end{abstract}




\section{Introduction}

Stars with initial masses $\gtrsim\,8\,$M$_\odot$ end their lives as core collapse supernovae \citep[CCSNe, e.g.][]{2004Eldridge, 2004Podsiadlowski, 2007Eldridge, 2012Janka}. The exact subtype and observable properties of a CCSN depend on the specific properties of the progenitor star and local environment. Whilst there are various progenitor models to explain the wide variety of CCSNe types, only a small fraction of progenitors have actually been observed to provide direct constraints \citep[e.g.][]{2009Smartt, 2015Smartt, 2017VanDyk, 2019ONeil}. As expected, the majority of these observations have been of the progenitors of the more common classes of CCSNe. As a result, when investigating rarer classes, indirect constraints on their progenitors from the SNe properties must be used -- typically in conjunction with stellar evolution models \citep[eg.][]{2024Tomoya, 2025Avishai,2025Zapartas}. \\

One subtype of SN with numerous questions still surrounding their progenitors are interacting SNe. These are SNe whose observed properties show the presence of interaction between their ejecta and surrounding circumstellar material (CSM). The most common type of interacting SNe are Type IIn \citep[SNe IIn, eg.][]{1990Schlegel, 1997Filippenko}. SNe IIn show narrow hydrogen emission-line features in their spectra, which are thought to arise from this ejecta interaction. Less common interacting SNe are Type Ibn \citep[SNe Ibn, e.g.][]{2007Foley, 2007Pastorello} and Type Icn \citep[SNe Icn, e.g.][]{2021Fraser, 2022Gal-Yam, 2022Perley}, each with no strong hydrogen features present in observed spectra. SNe Ibn show narrow helium emission-line features, SNe Icn lack any strong helium features and have narrow carbon lines, and SNe Ien that lack both strong helium and carbon features and have narrow silicon lines \citep{2025Schulze}. The absence of strong features in their spectra implies that the progenitors of these SNe have been completely stripped of their outer hydrogen, and in the case of SNe Icn/Ien their helium envelopes. This stripping either persisted until the end of the progenitor’s life, occurred shortly before explosion, or there was a mechanism that collected the lost mass close to the star, resulting in it being surrounded by dense CSM.\\

There have been observations of Type IIn SNe progenitors that suggest these events are from massive blue stars that retain their hydrogen envelope until shortly before explosion \citep{2007GalYam, 2009GalYam, 2011Smith}. So far, there have been no constraining limits or detections of the progenitors of SNe Ibn/Icn, described henceforth collectively as stripped envelope interacting SNe (SEISNe). The majority of SNe Ibn are associated with recent massive star formation \citep{2015Pastorello}, although a few notable exceptions have been found in regions with little to no star formation \citep{2013Sanders, 2025Warwick, 2025Dong}. This leads to the suggestion that the main route for SNe Ibn is the core collapse of a massive star, with a small fraction arising from a different evolutionary pathway \citep{2019Griffin}. The original model for the progenitors of Ibn SNe was massive Wolf-Rayet (WR) stars \citep{2007Pastorello, 2008Tominaga}. In this model the CSM is produced from the high wind-driven or eruptive mass loss in the final stages of their life. The contribution by low final mass ($\lesssim3\,$M$_{\odot}$) helium stars to the SN Ibn population has also been explored \citep[e.g.][]{2020Ning-Chen,2022Dessart,2025Ercolino,2025Takatoshi}. In this scenario binary interactions between the progenitor and its companion star are the cause of the stripping of its outer envelope and the formation of the dense CSM. \\

The ability to constrain the progenitors of SNe Ibn through statistical investigations is possible thanks to the sample of 71 events that have been classified as Ibn\footnotemark. For SNe Icn, their extreme rarity, with only 7 having been classified, leaves a limited sample to investigate their progenitor channels statistically. However, the similar observed properties of SEISNe; their fast rise times, similar peak luminosities and late time spectra, suggest that their progenitors may be similar \citep{2022GalYam, 2022Perly}. There additionally exists objects that are originally classified as Icn which then have helium features observed at later times causing them to appear as Ibn \citep[e.g.][]{2023Davis, 2023Pursiainen, 2025Gangopadhyay}. These transitional types suggest that SEISNe may exist as a continuum of objects arising from similar progenitors that have experienced different levels of stripping \citep{2023Pursiainen, 2025Gangopadhyay}.\\

\footnotetext{Based on a Transient Name Server, https://www.wis-tns.org/, query on 15/04/2026}


Research has been carried out simulating progenitor channels of SEISNe \citep{2025Avishai, 2025Ercolino}. The computational expense for carrying out these simulations is non-trivial, with each increase in parameters simulated increasing the total computation time by factors more. Modelling binary evolution leads to a considerable increase in the volume of parameter space that must be covered. Accounting for a range of initial orbits and companion mass ratios increases the dimensionality of the problem, exacerbating the computational demands. A way we can investigate a wide range of progenitor parameters is to make use of large stellar evolution grids created for the purpose of modelling the spectral energy distributions (SEDs) of galaxies, a technique known as stellar population synthesis \citep[SPS,][]{1991Charlot,1993Bruzual,2013Conroy}. \\

In this paper we make use of the Binary Population and Spectral Synthesis models \citep[BPASS,][]{2009Eldridge, 2012Eldridge, 2018Stanway, 2022Byrne} to investigate the potential progenitors of SEISNe and their surrounding CSM. We will test if the mass loss from both winds and Roche lobe overflow (RLOF) detailed in the BPASS models is sufficient to power, through interaction between the SN ejecta and the CSM, the luminosities observed in SEISNe. For those models that do match the luminosities, we were able to investigate the likelihood of the progenitor pathway arising and the environment that would produce them. We were also able to investigate the radio detectability of emissions from the interaction between the SN blast wave and CSM, and compare this to observations.\\

This paper is structured as follows: Section \ref{Meth} details the criteria used when selecting models from BPASS, the methods used in deriving the state of the models' CSM at explosion and for calculating a simple interaction powered light curve from our CSM models. Section \ref{Disc} discusses our results, \ref{Caveats} our modelling assumptions, and \ref{Con} summarises our findings. \\

\section{Methods}
\label{Meth}

\subsection{Introduction to BPASS}\label{BPASS}

The Binary Population and Spectral Synthesis Framework \citep{2009Eldridge, 2012Eldridge, 2017Eldridge, 2018Stanway, 2022Byrne} is a suite of models which enable the study of simple stellar populations (i.e. those formed in a single star formation event) at ages extending from 1\,Myr to 100\,Gyr. The framework consists of three core elements: a suite of custom, detailed binary stellar evolution models, a population synthesis formalism for combining these with weightings derived from an initial mass function and initial binary parameters, and a spectral synthesis formalism which assigns each star an observed spectrum, permitting the simulation of integrated light from an unresolved stellar population. Here we make use of the extensive grid of stellar evolution models, and the population weightings, making no use of the observed optical emission spectra.\\

The BPASS stellar model code is an offshoot of the Cambridge STARS code \citep{1971Eggleton, 1972Eggleton, 1973aEggleton, 1973bEggleton}, originally optimised for the calculation of supernova progenitors in binary systems \citep{2008Eldridge}. It solves for the energy generation and transport, composition, opacity, matter transport and hence detailed 1D stellar structure, at variable time steps which adapt to account for the rate of change of key variables. In addition, it produces detailed models of interacting binaries through a two step process. In the first, the initially more massive (primary) star is modelled in detail, while the slower-evolving secondary star is approximated using the rapid stellar evolution formalism of \cite{2000Hurley}. Roche Lobe overflow is calculated as described below, and its impact on subsequent evolution accounted for. In the second step, after the primary has reached the end of its life, the secondary evolution is calculated using the same detailed code, allowing the continued evolution of the binary, potentially including a remnant of the primary, to be determined.  The mass of any such remnant is determined from the stellar structure, and is defined as the mass which cannot be unbound by an injection of kinetic energy equivalent to $10^{51}$\,ergs. The possible outcomes of a supernova kick (unbinding the system, resulting in a merger or modifying orbital parameters) are accounted for by distributing the initial weighting of each primary model between a range of regularly-gridded secondary models at the population synthesis stage.\\

The detailed physics of the BPASS stars code is laid out in \cite{2017Eldridge}, with additional physics for low mass stars (not relevant here), further described in \cite{2018Stanway}. The code has been extensively validated and tested in a series of papers \citep[e.g. ][]{2016Wilkins, 2016Wofford}, and is widely used in the supernova progenitor and massive star communities \citep[e.g. ][]{2022Brennan, 2022Briel, 2024Briel, 2021Stevance, 2020Chrimes, Xiao_2015, Xiao_2018}. The models  used in this work comprise the BPASS v2.2 stellar model grid. This includes stellar evolution models at 13 metallicities ranging from Z=$10^{-5}$ to 0.040, where Z=0.020 is identified as Z$_{\odot}$. At each metallicity, single star and primary star models are calculated for stars with initial masses in the range 0.1 to 300 M$_{\odot}$. Initial binary mass ratios are calculated in the range q=M2/M1 = 0.1 to 0.9 at intervals of 0.1. Initial periods range from log(P/days)=0.0 to 4.0 at intervals of 0.2 dex.\\  



These properties make the BPASS v2.2 grid of stellar models well suited to investigating the progenitors of CCSNe, whose progenitors are young massive stars extending to low metallicities. The initial mass function used in BPASS is that of \cite{2001Kroupa} which is abroken power-law of the form $M^{\alpha}$ with a power-law slope $\alpha=-1.30$ for stars with masses from 0.1 to $0.5\,$M$_{\odot}$ and  a slope $\alpha=-2.35$ for masses above this. BPASS applies this IMF up to a maximum stellar mass of $300\,$M$_{\odot}$ alongside the empirically-determined initial binary parameter distributions from \cite{2017Moe}, which were originally derived at near-Solar metallicities, and are applied uniformly at all metallicities. These yield an effective binary fraction for massive stars (M>10M$_{\odot}$) of unity, and a near-constant distribution in initial log Period. The impact on populations of varying the binary parameters was explored in \citet{2020Stanway}. Effectively, approximately 70 per cent of massive stars are likely to interact with their companions at some point in their evolution \citep{2019Stanway}. This is consistent with empirical estimates \citep{2012Sana, 2014Sana}.\\

As described in Section 2.1.2 of \citet{2017Eldridge}, the mass-loss rate due to Roche lobe overflow (RLOF) is determined as \\
\begin{equation}
    \dot{M}_{R1}=F(M_1)\left[\ln\left(\frac{R_1}{R_{L1}}\right)\right]
\end{equation}
\\
\noindent where\\
\begin{equation}
    F(M_1) = 3\times10^{-6}[min(M_1,5.0)]^2
\end{equation}

\noindent and $R_{L1}$ is the Roche lobe radius, defined by \citet{1983Eggleton}. Material lost from the primary by Roche Lobe overflow is deemed to be accreted by the secondary unless it exceeds the Eddington limit for the secondary star, in which case it instead is ejected and contributes to the CSM. Super-Eddington accretion is permitted only for black-hole accretors. We ensure we take into account the mass accreted by the secondary star when modelling the CSM by tracking the mass of secondary star alongside the RLOF rate and comparing the two. For the models selected from BPASS we find a IMF weighted average for the accretion efficiency of the RLOF of 0.3 percent, and for those models that we determine to be SEISNe progenitors an efficiency of 4 percent.\\

Mass-loss rates are assigned based on surface properties and evolutionary state of the star. In the BPASS v2.2 models used in this work, the mass-loss rates of \citet{1988deJager} are applied to most Main Sequence stars, while the mass-loss rates of \citet{2001Vink} are used for OB stars (with effective temperatures above 12.5\,kK). For Wolf-Rayet stars (surface hydrogen below 40$\%$ and surface temperatures greater than 10 kK), the mass-loss rates of \citet{2000Nugis} are applied. Finally, for stars in the AGB regime, the mass-loss rates of \citet{2005Schroder} are used. Mass loss rates on the main sequence and giant branch scale as $Z'^{0.5}$ and those for OB stars as $Z'^{0.85}$ (Vink 2001) where $Z'=Z/0.020$.\\

\subsection{BPASS Model Selection}

For selecting stellar or binary models from the BPASS grid we made use of the HOKI package \citep{2020Heloise}. Using this package we selected models from BPASS v2.2 \citep{2018Stanway}, across all the metallicities present, based on criteria applied to the model parameters at their final evolutionary timestep. Our selection criteria are as follows:\\

\begin{enumerate}
    \item Carbon-Oxygen core mass $> 1.38\,$M$_{\odot}$
    \item Oxygen-Neon core mass $> 0.1\,$M$_{\odot}$
    \item Surface temperature log($T$/K) $\geq 4.45$
    \item Surface hydrogen mass fraction $X_\mathrm{surf} \leq 10^{-3}$
    \item Carbon-Oxygen core mass $< 16.2\,$M$_{\odot}$
\end{enumerate}

Criteria (i) and (ii) identify systems that will undergo core collapse, with criteria (i) ensuring the core is sufficiently massive \citep{1988Nomoto}, and criteria (ii) ensuring we only select objects that have finished core carbon burning at the end of their lives. Criteria (iii) and (iv) select those likely to end their lives in SN Ib/Ic \citep{2017Eldridge}, and criteria (v) removes any progenitors likely to result in a failed supernova. Criteria (iii) could exclude He-giants that also result in SNe Ib \citep[e.g.][]{2012Yoon}. However, within the BPASS models, any He-giants that could cool to below our threshold temperature would be in a binary system and therefore the interaction would prevent them from cooling. Criteria (v) comes from the result of \cite{2025Maltsev}, who set a lower threshold for this value (below which all SNe are successful) of $M_{CO} = 5.6\,$M$_{\odot}$ and upper threshold (above which all SNe fail i.e. a direct collapse) as $M_{CO} = 16.2\,$M$_{\odot}$. We use the upper threshold, and assume all the models whose progenitors are below this will result in a successful SNe. Whilst this will lead to higher SN rates from our simulations it means we are not removing any potential SEISNe progenitors. We do not apply any criteria requiring binary systems at this stage so that we can investigate the CSM from the single star models present in BPASS.\\

\begin{figure*}
    \centering
    \includegraphics[width=\textwidth]{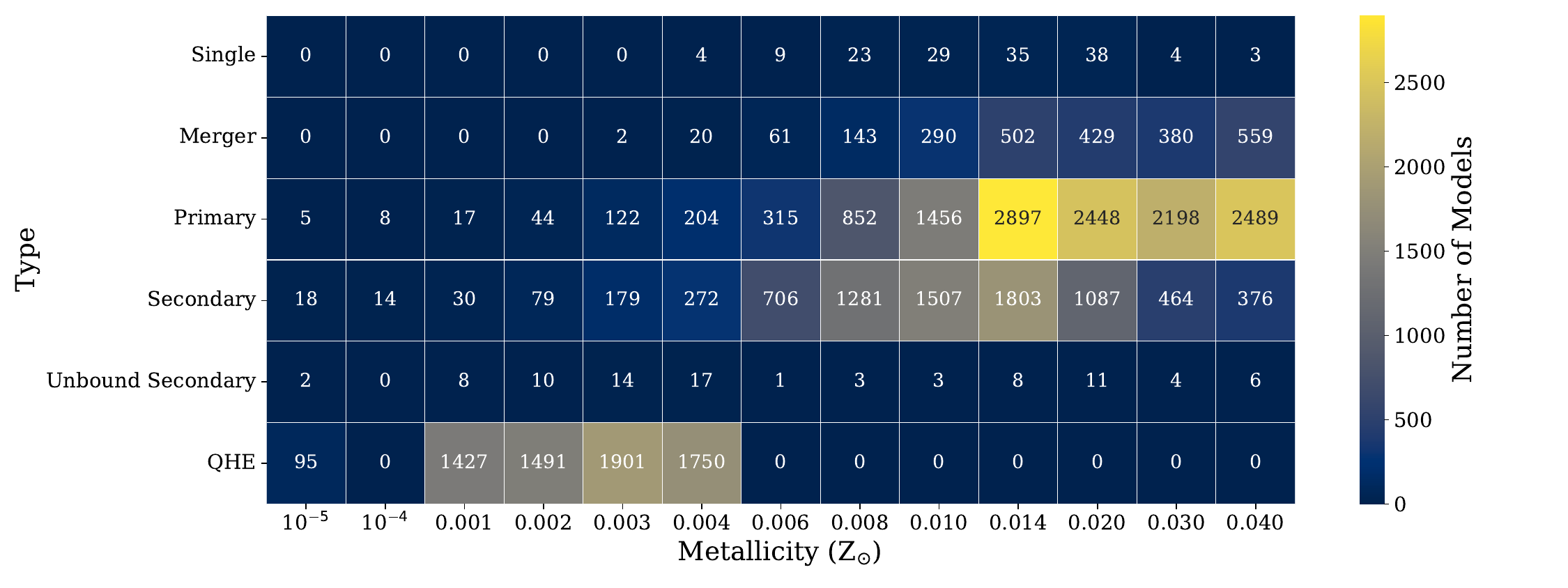}
    \caption{The distribution of the models that passed all of our selection criteria by metallicity and model type. A description of each model type is given in Table \ref{tab:stellar_types}.}
    \label{fig:Type_met}
\end{figure*}

Our selection criteria resulted in a final selection of 30,153 stellar evolution models. The breakdown of the model types, and number of each selected, is as detailed in Table \ref{tab:stellar_types}. These models were spread across 13 discrete metallicities from $Z=10^{-5}$ to $Z=0.040$. The full initial number of selected models by type and metallicity is detailed in Figure \ref{fig:Type_met}. In Figure \ref{fig:Bin_type} the IMF-weighted fraction of the selected models that undergo binary interactions is shown. Figure \ref{fig:Case_type} shows the types of mass transfer that the selected models undergo. The majority of the binary models with an initial primary star mass of $\leq70\,M_\odot$ undergo case B  mass transfer \footnote{The different types of mass transfer are described in more detail in Appendix A. Case B is mass transfer that starts after the donor leaves the main sequence but before helium ignition (during shell hydrogen burning).}, with all binary models $\leq30\,M_\odot$ undergoing it. In the last 100\,kyr of the models evolutions, those with a primary star that has a ZAMS mass of $\gtrsim 10\,$M$_\odot$ lose most of their mass as wind, and those with $\lesssim 10\,$M$_\odot$ as RLOF (Figure \ref{fig:WindRLOF}).\\ 

We show equivalent plots for only the subset of models that result in SEISNe in Figures  \ref{fig:Bin_type_SEISNe}, \ref{fig:Case_type_SEISNe}, and \ref{fig:WindRLOFSEISNe}. In Figure \ref{fig:Bin_type_SEISNe} the IMF-weighted fraction of the models we determine to be SEISNe that undergo binary interactions is shown, as well as the nature of that interaction. Figure \ref{fig:Case_type_SEISNe} shows the types of mass transfer that the models we determine to be SEISNe undergo. In the last 100\,kyr of the SEISNe models evolutions, only those with a primary star that has a ZAMS mass of $< 11\,$M$_\odot$ or $> 70\,$M$_\odot$ lose more mass to RLOF than through their winds (Figure \ref{fig:WindRLOFSEISNe}).\\

\begin{table*}
\centering
\caption{BPASS Model Types}
\begin{tabular}{c r p{9cm}}
\hline
Type & Number & Description \\
\hline
$-1$ & 145 &
Single stars that evolved without any binary interaction. \\
0 & 2,386 &
Single stars formed from mergers on or before the Main Sequence. \\
1 & 13,055 &
Primary stars in binary systems. \\
2 & 7,816 &
Secondary stars in binaries after the full evolution of the primary star through to a compact object. \\
3 & 87 &
Secondary stars ejected from a binary due to a supernova kick from the primary. \\
4 & 6,664 &
Secondary stars that were spun up during the evolution of the primary star, became fully mixed, and experienced quasi-chemically homogeneous evolution (QHE). \\
\hline
\end{tabular}
\label{tab:stellar_types}
\end{table*}

\subsection{CSM Modelling}
For investigating the CSM of each model at core collapse we consider the last 100\,kyrs of each model's mass loss history. We considered the mass loss both due to stellar winds and RLOF. The time frame was chosen as it is the expected upper limit for the lifetime of any potential circum-binary disk (CBD) that could form \citep{2023Tuna}. Wind-mass loss from earlier times will have moved out to large radii at this time and have become tenuous. Therefore it won't affect the early light curve or spectroscopic behaviour of the SEISNe. We select 3 different scenarios for how the mass loss is distributed around the progenitor(s):
\begin{enumerate}
    \item \emph{Pure Wind}: All the lost mass is distributed in the form of a (spherically symmetric) wind. This scenario presents an upper-limit on the density of the spherically-distributed material. This scenario includes both the single and binary models from those that passed our initial selection criteria.
    \item \emph{Wind Capture}: The mass lost through RLOF is distributed in a CBD, while the wind mass loss is ejected evenly in all directions. Any wind mass loss within the opening angle of the disk is collected and distributed evenly inside the disk. This provides an upper limit on the mass of the CBD assuming all RLOF and intersecting wind mass-loss populates the disk. This scenario includes only binary models from those that passed our initial selection criteria.
    \item \emph{Free Streaming}: As with the \emph{wind capture} scenario, the mass lost through RLOF is distributed in a CBD and the wind is ejected in all directions. In this case we assume that the wind passes freely through the disk and does not accumulate, thus giving a lower limit on the mass of the CBD assuming all RLOF mass-loss populates the disk. This scenario includes only binary models from those that passed our initial selection criteria.
\end{enumerate}

\begin{figure*}
    \centering
    \includegraphics[width=\textwidth,height=0.9\textheight,keepaspectratio]{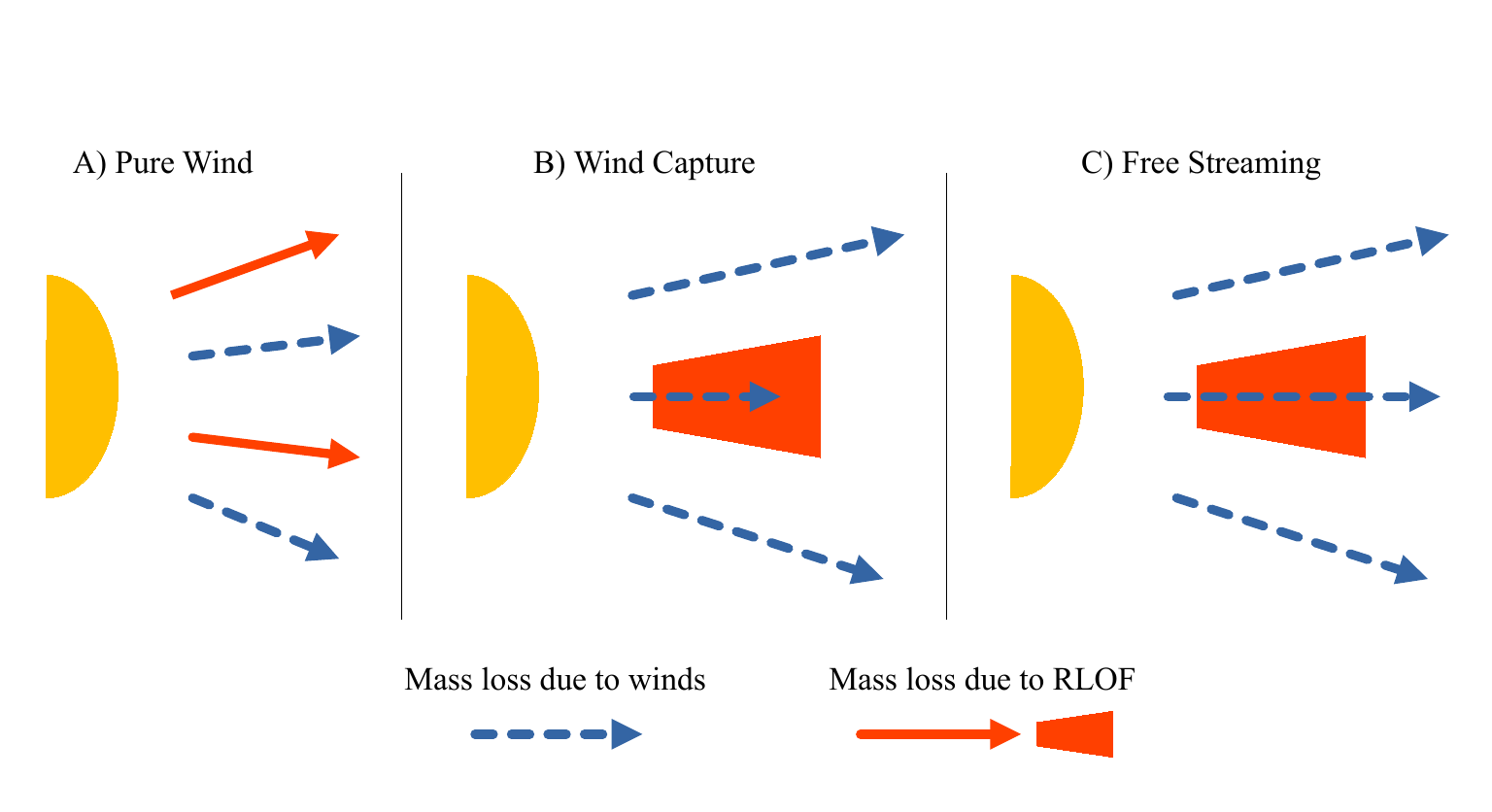}
    \caption{Diagram showing the three different scenarios considered in our CSM modelling. Left: Both RLOF and wind material are distributed as a wind. Middle: RLOF material forms a CBD that captures all the wind within the disk’s opening angle, representing an upper limit on the density of the CBD. Right: RLOF material forms a CBD while the wind streams freely through it, presenting a lower limit on the density of the CBD.}
    \label{fig:Diagram}
\end{figure*}

A diagram of the scenarios is presented in Figure \ref{fig:Diagram}. As the \emph{wind capture} and \emph{free streaming} scenarios require a CBD these scenarios only include binary models, where as the \emph{pure wind} scenario includes both binary and single star models from BPASS. The opening angle of the CBD can vary from $10^{\circ}$ to $30^{\circ}$ \citep{2025Scherbak}. As narrower disks would result in lower SN luminosities (as there will be less interaction between the ejecta and the disk), we choose the opening angle of the disk in both disk scenarios to be $10^{\circ}$ as a conservative value. When distributing the mass as wind we consider the time and velocity at which it was lost. This velocity follow \cite{2000Nugis} and is described by
\begin{equation}
\log\, v_{\infty}/v_{\rm esc}({\rm core}) = 0.61 - 0.13\log L+0.30\log Y
\label{WN}
\end{equation}
for WN stars, and
\begin{equation}
\log\, v_{\infty}/v_{\rm esc}({\rm core}) = -2.37 + 0.43 \log L-0.07 \log Z
\label{WC}
\end{equation}
for WC stars. $v_{\infty}$ is the terminal velocity of the wind, $v_{\rm esc}({\rm core})$ is the escape velocity of the core, $L$ the stars luminosity in $L_{\odot}$, and $Y$ and $Z$ are the He and metal abundances, respectively. WN and WC stars are WR stars whose spectra are dominated by nitrogen and carbon features, respectively and we select the appropriate velocity above based on a surface abundance threshold of $(C + O)/He = 0.03$ and the core mass from the BPASS models. These velocities are only appropriate for WR stars, whereas many of our models end as helium giants. In these cases the mass loss rates from BPASS are also likely not appropriate, with \cite{2020Andreas} showing the mass-loss rates from less massive helium stars are lower. This difference in mass-loss does not influence the conclusions drawn from our modelling, as detailed in section \ref{Caveats}.

In the models where there is a CBD, we set the density of the disk to be constant and to have an inner radius of $3a$ (where $a$ is the semi-major axis of the binary), which is caused by eccentricity resonances \citep[per][]{1994Artymowicz}. We set the outer radius of the disk as $100a$, as this is comparable to the radius at which photo-evaporation becomes important, see equation 29 of \cite{2023Tuna}.\\

The relevant model parameters we have from BPASS are as follows:\\

\begin{enumerate}
    \item $M_1$: the mass of the primary star (primary star is defined as the initially more massive star)
    \item $M_2$, the mass of the secondary star
    \item $\dot{M}_{1, \rm{wind}}$, the mass loss rate due to winds of the primary star
    \item $\dot{M}_{2, \rm{wind}}$, the mass loss rate due to winds of the secondary star
    \item $\dot{M}_{RLOF}$, the rate of mass loss (or gain) by both stars due to RLOF 
    \item The surface composition of the primary and secondary star at discrete times throughout the evolution of the star
\end{enumerate}

\begin{figure*}
    \centering
    \includegraphics[width=\textwidth]{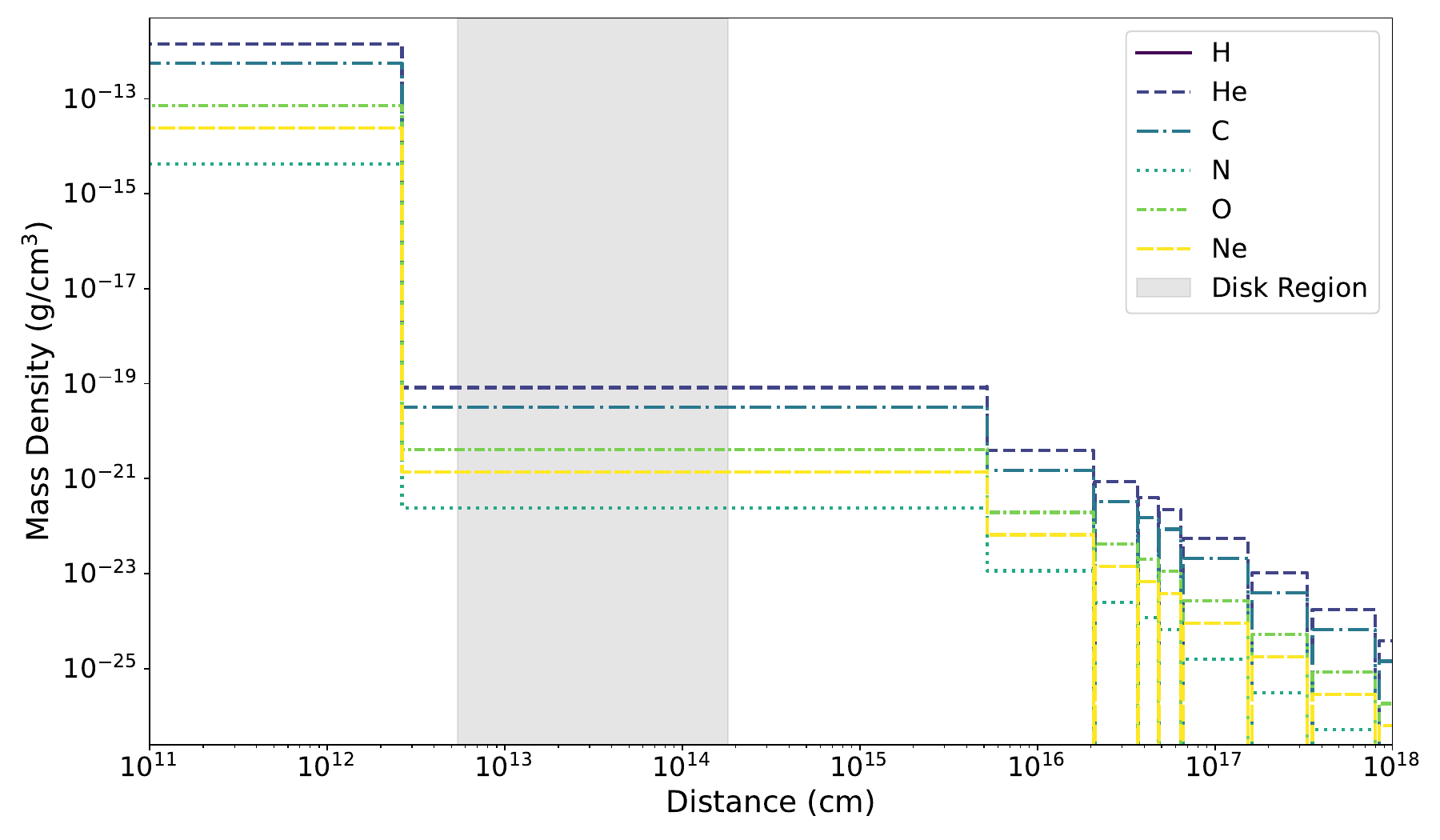}
    \caption{Example CSM density profile derived from a progenitor model with an initial mass of the primary star of 30\,M$_{\odot}$, a secondary star with a mass of 9\,M$_{\odot}$, an initial separation of 35.86\,R$_{\odot}$, with a metallicity of $z = 0.014$. The lines represent the total mass density as a function of distance from the progenitor, decomposed by element. The shaded region shows the radius of the CBD for this progenitor model.}
    \label{fig:Examp}
\end{figure*}

From these we can obtain the total mass lost due to winds and RLOF, the composition of that mass loss, and, for the wind, the velocity. We assume that the mass loss rates and velocity are constant over each model time step. This allows us to create CSM models comprising of shells of constant density and moving with a constant velocity at the time of the SN for each progenitor model. An example of these CSM models is shown in Figure \ref{fig:Examp}. Each individual shell in the final CSM model is created from the mass-loss of one or more distinct time steps. When it is formed from more than one time step this will be due to material released at later times catching up to earlier material due to the wind velocity of the star increasing as it evolves. When material catches up to other material, the overlapping region between the two will be split off as a separate shell by averaging the velocity and summing the density of them.\\

The results of our CSM modelling finds only 1812 out of our 30,153 selected models have $>1$ per cent of their CSM comprised of H, with the most extreme case having a CSM that was comprised of $37$ per cent H. Of the 2006 we later determined to be SEISNe only 67 have $>1$ per cent of their CSM comprised of H, with the most extreme case having a CSM that was comprised of $14$ per cent H. We therefore assume for the rest of out modelling that the CSM is He dominated and that the resulting SNe would lack strong H features in their spectra.

\subsection{Interaction Light curve}
When investigating the interaction powered light curve from the resulting SNe in our combined BPASS and CSM models we considered the SN's ejecta along with the material it sweeps up from the CSM, which we refer to as the cold dense shell \citep[CDS,][]{2001Chugai}. For the ejecta mass we used the value from BPASS for SNe with a total kinetic energy of $E_{SN} = 10^{51}$\,ergs, wherein the ejecta mass is obtained by integrating the binding energy of the stellar structure towards the core until it reaches this fiducial kinetic energy \citep{2004Eldridge}. Assuming all the kinetic energy of the SN goes into the kinetic energy of the ejecta we calculated the ejecta velocity in each case according to $v_\mathrm{ejecta} = (2E_\mathrm{SN}/M_\mathrm{ejecta})^{1/2}$. With these properties of the ejecta and our model's CSMs we then made a light curve model for the resulting SN so we could compare key parameters of its evolution, rise time and peak luminosity to observations. In order to model these light curves we assumed that the observed luminosity arises from the conversion of the kinetic energy lost by the CDS into radiation, with an efficiency of 100 per cent, as the CDS propagates outward. Given these assumptions the average luminosity, $L_{avg, shell}$, for the interaction with a shell of CSM with a constant density is given as,
\begin{equation}\label{Lum_eq}
L_{avg, shell} = \frac{1}{4dr}M_{0, shell}v_{0, shell}^3\left[1-\frac{M_{0, shell}^2}{(M_{0, shell}+dm)^2}\right]
\end{equation}
where $dr$, is the width of the shell, $M_{0, shell}$ and $v_{0, shell}$ are the initial mass, and velocity (relative to the CSM), respectively, of the CDS as it encounters the shell, and $dm$ is the total mass of the shell. A full derivation of equation \ref{Lum_eq} is presented in Appendix \ref{Equ}. Within each individual model $L_{avg, shell}$ is calculated for each distinct shell of CSM, with $M_{0, shell}$ and $v_{0, shell}$ being updated along side. We use the midpoint of the density shells for the peak time and the half width of the shells as the error in this peak time. As the luminosity is only an average across the shell the actual peak expected from our models could be greater, in the extreme cases up to a factor of 0.5 greater. We use the same method for calculating interaction luminosity from both the wind and disk CSM. When applying it to one of the two disk scenarios we only calculate the luminosity for the disk and wind in the opening angle of the disk. This is then combined with a fraction\footnote{Here, the fraction refers to the component of the wind luminosity arising from regions not covered by the disc opening angle.} of the luminosity from the \emph{pure wind} scenario to get the full luminosity of the relevant disk scenario.\\

In addition to the luminosity from interaction power we also include in our model light curves the luminosity due to radioactive decay from $^{56}$Ni. We use the radioactive decay models of \cite{2012Chatzopoulos} and calculate the gamma-ray optical depth of the ejecta as $\tau_{\gamma} = 3\kappa_{\gamma}M_{ej}/4{\pi}R^{2}$, with $\kappa_{\gamma} = 0.03\,$cm$^2$g$^{-1}$ \citep{1995Swartz}. In each model we assume the ratio of $^{56}$Ni mass to ejecta mass of 1:14 \citep{2016Lyman}. This ratio is derived from Type Ib/Ic SNe, however previous work has demonstrated SNe Ibn/Icn produce noticeably less $^{56}$Ni \citep{2016Moriya, 2022Perley}. Due to the fast declining nature of SEISNe light curves obtaining accurate $^{56}$Ni ratios is difficult so we use the ratio for Type Ib/Ic SNe in this work. We combined the $^{56}$Ni with our interaction luminosity to determine a total luminosity for each time step. \\

\subsection{Synchrotron Emission Detectability}
We wanted to determine whether it would be possible to detect radio emission from the interaction of the resulting SN's blast wave with the CSM in our models. To do this we used the analytical expressions from \cite{1998Chevalier} to determine the spectrum of the emission. We then tested whether the synchrotron emission produced by the interaction of the blast wave with the disk can be detected over the frequency range 400 MHz to 869 GHz by the Giant Metrewave Radio Telescope \citep[GMRT,]{2019Patra}, MeerKAT \citep{2016Jonas}, the Karl G. Jansky Very Large Array (VLA) and the Atacama Large Millimeter/submillimeter Array (ALMA). In the case that more than one telescope can observe in a given frequency range, we have used the detection limits from the most sensitive telescope. The detection thresholds were calculated for a 3$\sigma$ detection and assume 1 hour on target (excluding calibration scans) with dual polarizations. In each case we have taken the recommended loss of bandwidth due to radio frequency interference into account, and have assumed the minimum recommended number of antennas on sky. The detection limits were estimated using the telescopes' online exposure calculators\footnote{These are available at https://apps.sarao.ac.za/calculators/continuum (MeerKAT), http://www.ncra.tifr.res.in/~secr-ops/etc/etc.html (GMRT), https://obs.vla.nrao.edu/ect/ (VLA) and https://almascience.eso.org/proposing/sensitivity-calculator (ALMA)} and are listed in Table \ref{tab:radiolimits}. \\

\begin{table*}
\centering
\caption{3$\sigma$ detection limits for one hour on target}
\label{tab:radiolimits}
\begin{tabular}{cccc}
\hline
Central Frequency (GHz) & 3$\sigma$ Detection Limit ($\mu$Jy/beam) & Telescope & Assumptions \\
\hline
0.4   & 57   & GMRT   & 120\,MHz usable bandwidth with a fudge factor of 3 and naturally weighted images \\
0.65  & 50   & GMRT   & 200\,MHz usable bandwidth with a fudge factor of 3 and naturally weighted images \\
0.816 & 40   & MeerKAT & Robust$=0$ image weighting, 544\,MHz of usable bandwidth, and no UV tapering \\
1.4   & 32   & VLA    & A configuration, robust image weighting and 0.6\,GHz of usable bandwidth \\
3.0   & 18   & VLA    & A configuration, robust image weighting and 1.5\,GHz of usable bandwidth \\
6.0   & 10   & VLA    & A configuration, robust image weighting and 3.4\,GHz of usable bandwidth \\
10.0  & 9    & VLA    & A configuration, robust image weighting and 3.4\,GHz of usable bandwidth \\
15.0  & 9    & VLA    & A configuration, robust image weighting and 5.28\,GHz of usable bandwidth \\
22.0  & 21   & VLA    & A configuration, robust image weighting and 5.28\,GHz of usable bandwidth \\
33.0  & 18   & VLA    & A configuration, robust image weighting and 7.5\,GHz of usable bandwidth \\
42.5  & 24   & ALMA   & Only uses the 12\,m array, with 7.5\,GHz of usable bandwidth \\
100.0 & 32   & ALMA   & Only uses the 12\,m array, with 7.5\,GHz of usable bandwidth \\
144.0 & 34   & ALMA   & Only uses the 12\,m array, with 7.5\,GHz of usable bandwidth \\
184.0 & 230  & ALMA   & Only uses the 12\,m array, with 7.5\,GHz of usable bandwidth \\
243.0 & 47   & ALMA   & Only uses the 12\,m array, with 7.5\,GHz of usable bandwidth \\
324.0 & 210  & ALMA   & Only uses the 12\,m array, with 7.5\,GHz of usable bandwidth \\
442.0 & 390  & ALMA   & Only uses the 12\,m array, with 7.5\,GHz of usable bandwidth \\
661.0 & 530  & ALMA   & Only uses the 12\,m array, with 15\,GHz of usable bandwidth \\
869.0 & 1300 & ALMA   & Only uses the 12\,m array, with 15\,GHz of usable bandwidth \\
\hline
\end{tabular}
\end{table*}

The radio emission occurs due the interaction between the SN's blast wave and the surrounding CSM producing synchrotron emission from the forward shock. That emission is attenuated by synchrotron self-absorption and free-free absorption. We calculate this emission following the formalism by \cite{1998Chevalier}, and using equations from \cite{2022DeMarchi} and \cite{2025Sfaradi}. To get the values of the synchrotron emission from our models we take the median disk density from our models in the \emph{free streaming} scenario, to obtain a lower limit on the CBD density, of $1.2\times10^{-13}\,$g$\,$cm$^{3}$. We then investigate the spectral peak flux density $F_{\nu_a}$, and frequency $\nu_a$ of the emission at radii out to the the greatest extent of the disks in our models, $4.0\times10^{15}\,$cm. This was done with the following equations:\\

\begin{equation}
F_{\nu_a} \;=\;
\left[
\frac{
B^{\,2p+13}\;
R^{\,2p+13}\;
c_5^{\,p+4}\;
\left(\dfrac{\epsilon_e}{\epsilon_B}\right)^{5}\;
f^{5}\;
(p-2)^{5}\;
E_1^{\,5(p-2)}
}{
6^{5}\;
\pi^{\,1-p}\;
D^{\,2p+8}\;
c_6^{\,p-1}
}
\right]^{\frac{1}{p+4}}
\end{equation}

\begin{equation}
\nu_a = 2 c_1 B \, 
\Bigg[ \frac{36 \pi^3 c_5}{\alpha^2 f^2 (p-2)^2 c_6^3 E_1^{2(p-2)} D^2 F_\nu} \Bigg]^{-2/(2p+13)}
\end{equation}\\

where $R$ is the radius of the radio emitting shell, $\epsilon_e$ is the fraction of the shock wave energy deposited in relativistic electrons, $\epsilon_B$ is the fraction of the shock wave energy converted to magnetic field, $f$ is the emission filling factor (the filling factor sets the size of the radio emitting region, see \cite{1998Chevalier} for further details), $E_1$ is the electron rest mass energy, and $D$ is the luminosity distance to the SN. The constants $c_1$, $c_5$, $c_6$ can be found in \cite{1970Pacholczyk}. We assume $p=3$ (as is typically observed for SNe at these evolutionary phases, e.g \cite{2022DeMarchi}), and equipartition conditions (therefore $\epsilon_e=\epsilon_B=0.1$). The emission filling factor $f$ is calculated using the following equation from \cite{2026Lazda}:\\

\begin{equation}
    f = (1 - \gamma^{3})sin\theta
\end{equation}\\

where $\theta$ is the opening angle of the CBD, and $\gamma=0.7$ from \cite{2026Lazda}. The magnetic field strength $B$ is calculated from:\\

\begin{equation}
    B = \sqrt{9\pi\epsilon_B\rho_\mathrm{CSM}v_\mathrm{sh}^2}
\end{equation}\\

where $\rho_\mathrm{CSM}$ is the density of the disk and $v_\mathrm{sh}$ is the velocity of the blast wave. We evaluate $F_{\nu_a}$ and $\nu$ at distances of 10, 50, 100 and 500\,Mpc.\\

After calculating $F_{\nu_a}$ and $\nu$ we then calculate the flux density, $F$, at the central wavelengths in Table \ref{tab:radiolimits} following:\\

\begin{equation}
    F = F_{\nu_a} \left(\left(\frac{\nu}{\nu_a}\right)^{\alpha_1/s}+\left(\frac{\nu}{\nu_a}\right)^{\alpha_2/s}\right)^{s}e^{-\tau}
\end{equation}\\

where $\alpha_1=5/2$, $\alpha_2=-1$, and $s=1$ as expected for a synchrotron self-absorption spectrum at these timescales post-explosion. The $e^{-\tau}$ term is to account for free-free absorption (FFA) where $\tau$ is the FFA optical depth given by \citep{1979Rybicki}:

\begin{equation}
    \tau = \int^{R_1}_{R_0}0.018Z^{2}n_en_iT_e^{-1.5}\nu^{-2}g_{ff}\,dr
\end{equation}\\

We use the Gaunt factor $g_{ff}=5$ (following the approach of \citealt{2025Nayana}), and assume a helium dominated CSM so that $Z=2$, $n_e=2n_i$, and $n_i=\rho_\mathrm{CSM}/m_{\alpha}$ where $m_{\alpha}$ is the mass of the helium nucleus. We set $T_e=10^{5}\,$K (following the approach of \citealt{2025Sfaradi}).

\subsection{Rates}

To calculate the expected observed SNe rates from the selected BPASS models we start from the model IMF. This model IMF specifies the mass expected to populate each model per $10^{6}\,$M$_{\odot}$ of star formation. For the primary models (singles, merger, and binary) we know the initial mass of the models and so can work out the expected number of that configuration, and so the expected SNe rate. This rate needs to be adjusted to take into account the star formation history, which itself needs to be adjusted according to the metallicity of the model. We use the star formation rate from \cite{2014Madau}, which is given by

\begin{equation}
\psi(z) = 0.015 \frac{(1 + z)^{2.7}}{1 + ((1 + z)/2.9)^{5.6}}\,\text{M}_{\odot}\,\text{yr}^{-1}\,\text{Mpc}^{-3}
\end{equation}
\\

where $z$ is redshift. This is combined with the metallicity scaling from \cite{Langer_2006}, using their fiducial values, given by

\begin{equation}
\Psi(\frac{Z}{Z_{\odot}}) = \frac{\hat{\Gamma}(\alpha + 2, (Z/Z_{\odot})^{\beta}10^{0.15\beta z})}{\Gamma(\alpha + 2)}
\end{equation}
\\

where $\hat{\Gamma}$ and $\Gamma$ are the incomplete and complete gamma function, $Z$ is metallicity, and $\alpha$ and $\beta$ are constants. This gives us a way of modifying our SNe rate taking into account the metallicity of the model and, using the age of the model at explosion, the redshift. \\

The expected occurrence of the different CSM profiles is quoted per $10^6\,$M$_\odot$ of star formation. This rate is computed for single, merger, and primary star model types, since for these the fraction of the $10^6\,$M$_\odot$ that exists in each configuration is known. As a result, their occurrence rates can be calculated directly. In contrast, the remaining models are secondary star evolution models. As explained in Section \ref{BPASS}, multiple primary star model evolutionary pathways converge onto a given secondary star model's evolution. Secondary models do not map directly to individual primaries, but instead are selected assuming a distribution of supernova kick outcomes.  This requires assumptions regarding the kick velocity distribution, including how it scales with black hole remnants, and the kick direction relative to the binary orbit. As such it introduces considerably more free parameters and the mass weighting assigned to each model does not directly map to a number of stars, and hence event rate. These secondary models account for only 2.2 per cent of the models (by number) that have SEISNe-like light curve parameters, and the secondary stars contribute only 0.5 per cent of the total mass per $10^6\,$M$_\odot$ of star formation. Therefore they contribute negligibly to the expected SN rates and are not discussed further.\\

\begin{figure*}
    \centering
    \includegraphics[width=\textwidth,height=0.9\textheight,keepaspectratio]{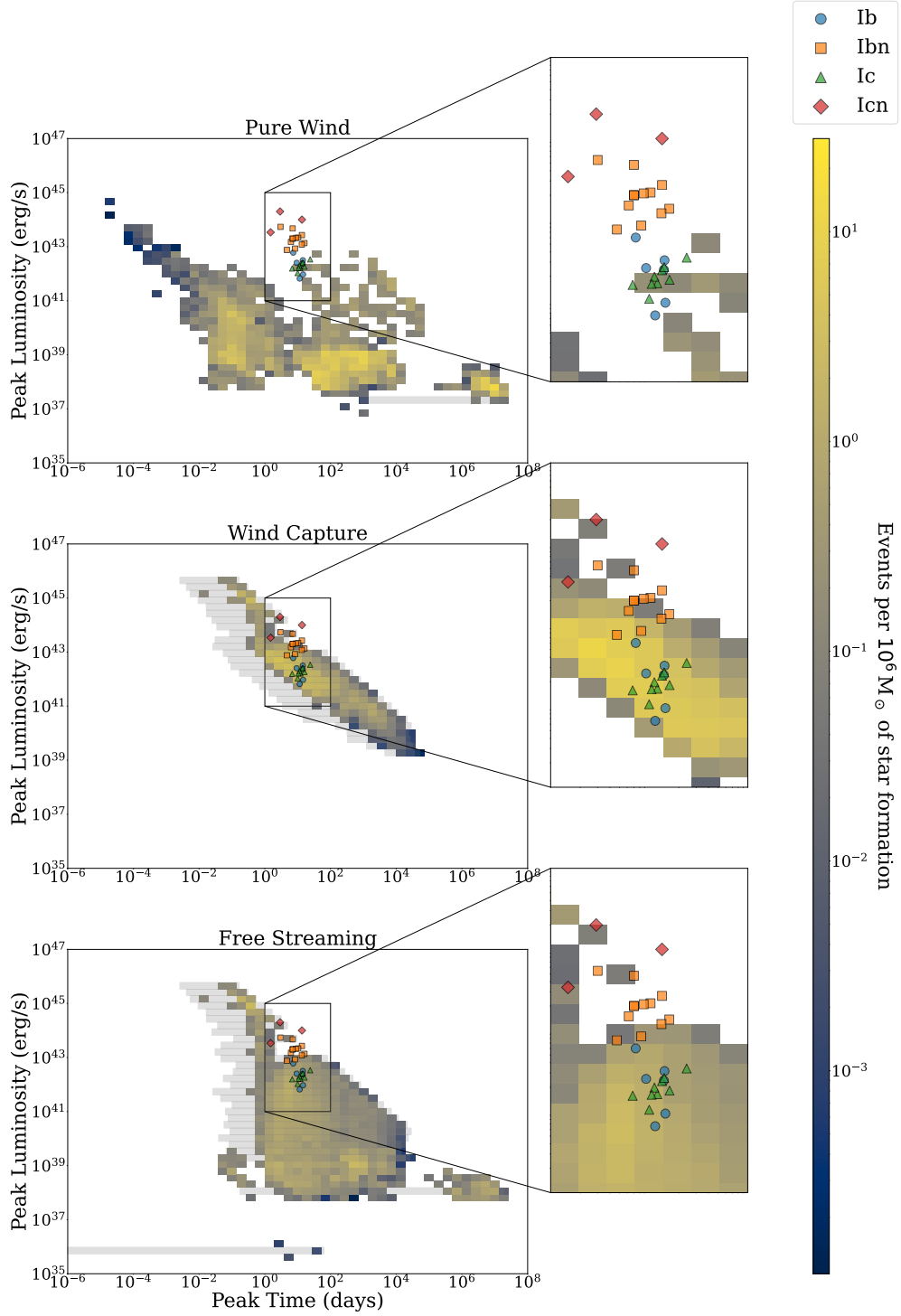}
    \caption{Heat map showing the range of peak values for modelled interaction light curves from the primary star BPASS models for our three scenarios, for the \emph{pure wind} (top), \emph{wind capture} (middle), and \emph{free streaming} (bottom) models. The heat map level shows the expected number of SNe with the peak parameters per $10^{6}\,$M$_{\odot}$. Peak luminosity is taken at the epoch of the peak interaction power including  $^{56}$Ni decay at this same epoch, and is not necessarily the peak of the SN's light curve. The transparent bars represent the uncertainty in the time of the light curve peak time. Overlaid are the observed peak properties of Type Ib (circles), Ic (triangles), Ibn \citep[squares,][]{2017Hosseinzadeh, 2023Pursiainen}, Icn \citep[diamonds,][]{2022Perly, 2022Gal-Yam, 2023Davis}.}
    \label{fig:Lum_rise}
\end{figure*}

\begin{figure*}
    \centering
    \includegraphics[width=\textwidth,height=0.9\textheight,keepaspectratio]{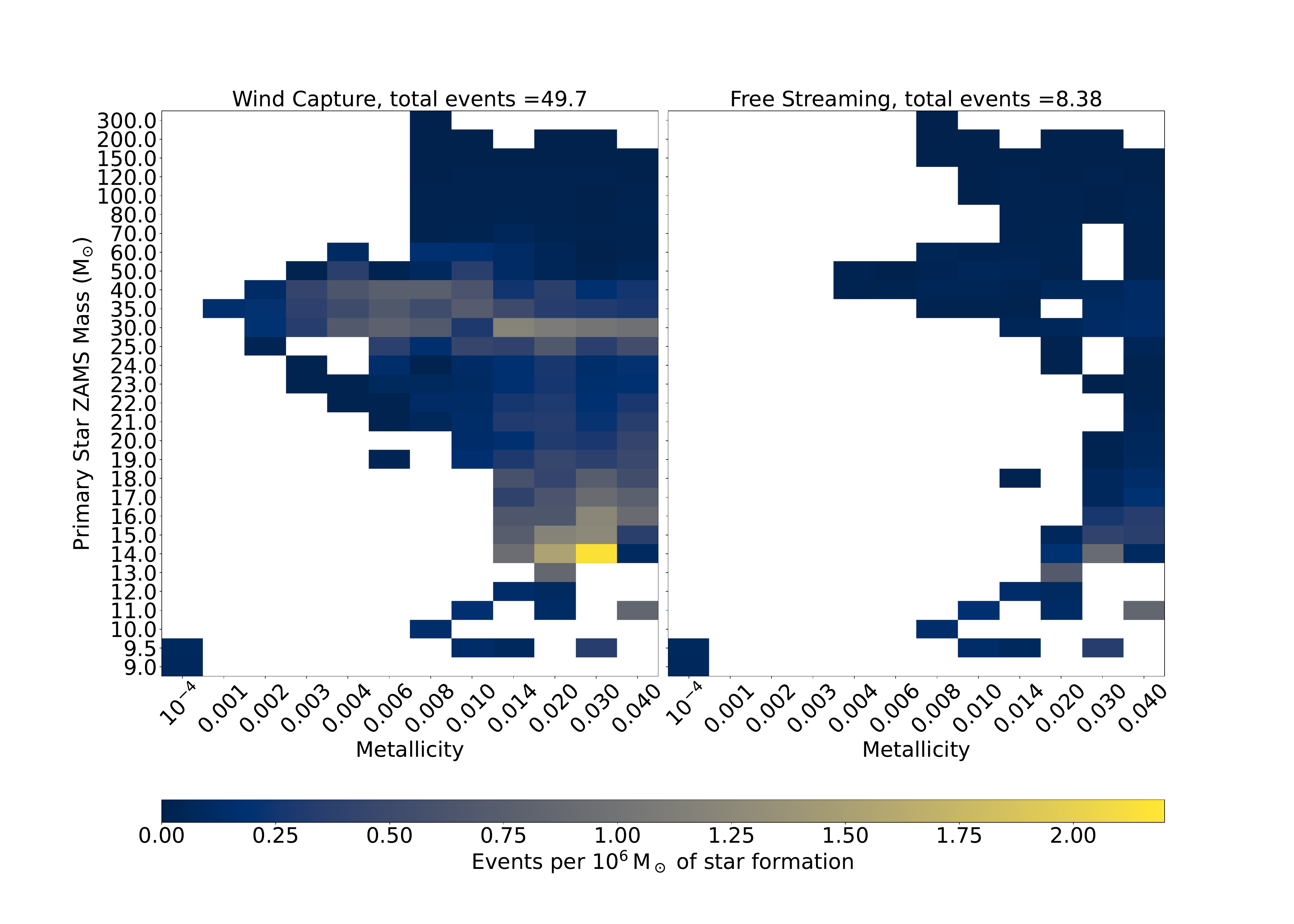}
    \caption{The initial ZAMS mass and metallicity of the primary star in the models that result in interaction powered light curves with peak parameters in the same range as those from observed SEISNe. This is shown for two of our three scenarios: \emph{wind capture} (left), and \emph{free streaming} (right) models. Only these two scenarios are shown as the \emph{free streaming} scenario could not produce peak parameters in the same range as those from observed SEISNe. The heat map level shows the expected number of SNe produced by progenitors with the corresponding ZAMS mass and metallicity per $10^{6}\,$M$_{\odot}$. The \emph{wind capture} model results in a greater expected number of events, with most occurring at high metallicities and from primary stars with high ZAMS masses.}
    \label{fig:Mass_met}
\end{figure*}

\begin{figure*}
    \centering
    \includegraphics[width=\textwidth,height=0.9\textheight,keepaspectratio]{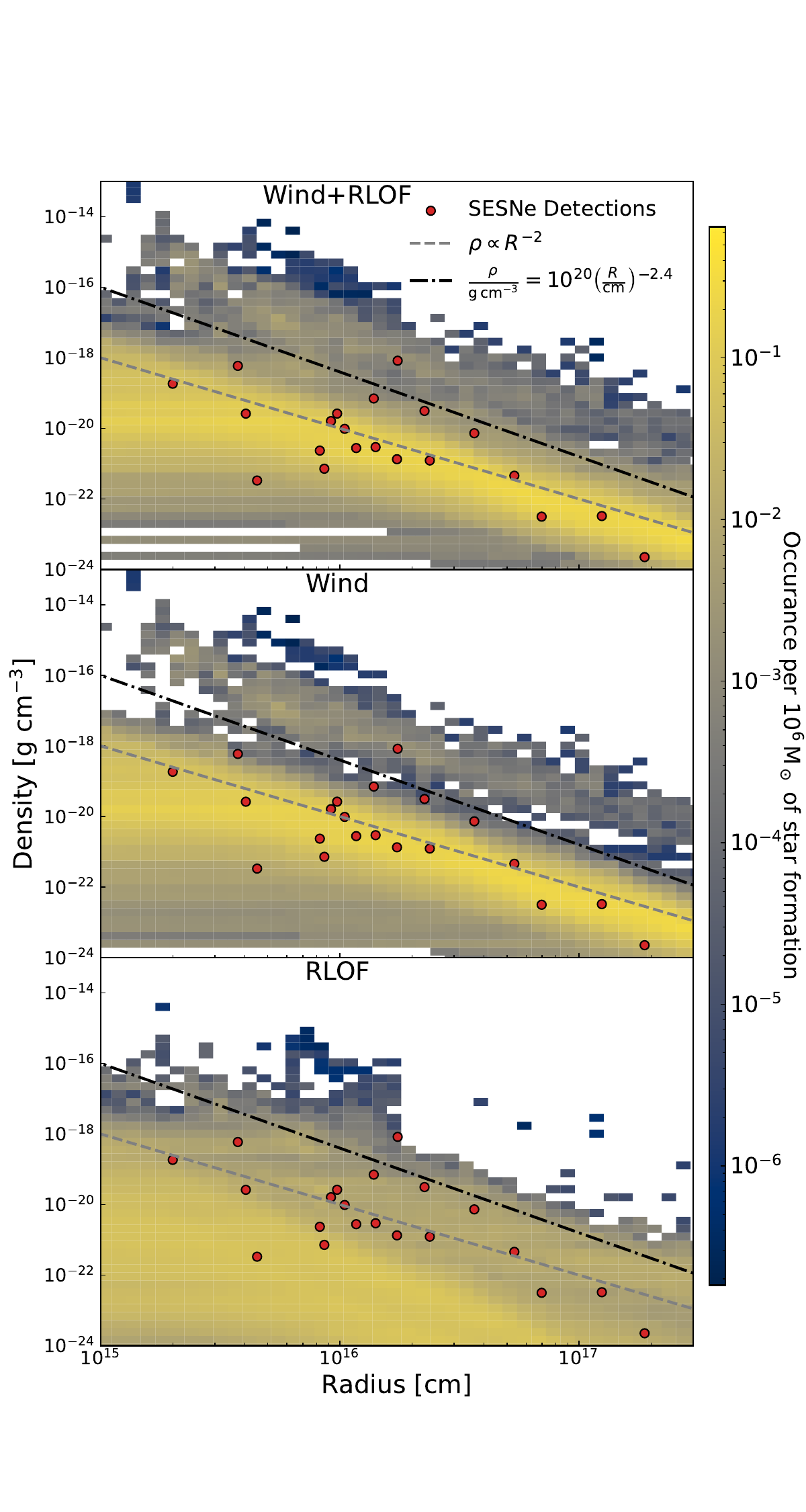}
    \caption{The heat map shows how many SN progenitors (per $10^6\text{M}_\odot)$ have a CSM profile with the given density at the given radius (just prior to explosion). This gives an indication of the expected CSM densities weighted across all our models. The rate is the expected number of objects in the bin per $10^6\,\text{M}_{\odot}$ of star formation. The observed densities of 22 Ibn SNe (derived from radio observations) from \citet{2025Sfaradi} have been overlaid. (Top) Shows the density for the combined wind and RLOF mass loss, (Middle) only the wind mass loss, and (Bottom) only the RLOF mass loss. The dashed line shows the line of $\rho \propto R^{-2}$, whilst the dot-dashed line shows $\frac{\rho}{\mathrm{g\,cm^{-3}}} = 10^{20} \left(\frac{R}{\mathrm{cm}}\right)^{-2.4}$. Above the dot-dashed lined all all the models that populate the region are merger models.}
    \label{fig:Density_dis}
\end{figure*}

\begin{figure*}
    \centering \includegraphics[width=\textwidth,height=0.9\textheight,keepaspectratio]{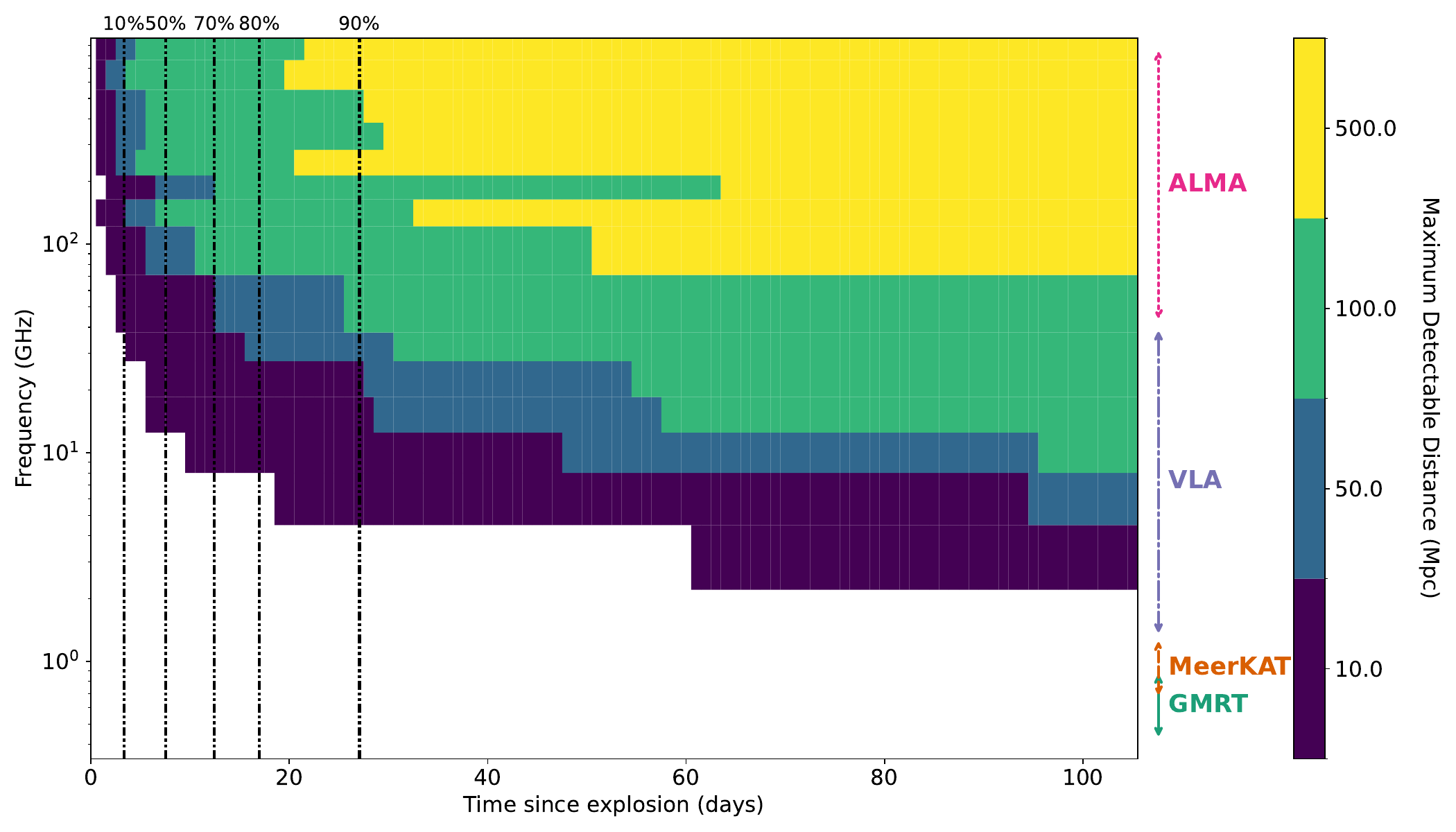}
    \caption{Prospects for a 3$\sigma$ detection in a 1 hour observation of the radio synchrotron emission from the interaction between the SN blast wave and a CBD with the median density of all our model CBDs. We invesigate up to the maximum disk size in our models of $10^{15}$\,cm. We account for synchrotron self-absorption, but not FFA. FFA is negligible outside the disk, so these results are vaild for observers with viewing angles outside the plane of the disc. For observes in the plane of the disc the FFA renders the disc interaction unobservable. The dotted vertical lines indicate the fraction of the CBDs from our models that would no longer be observable at a given time post explosion for a blast wave travelling at $4345\,\text{km}\,\text{s}^{-1}$. The different regions show the furthest distance that the emission could be detected at a given time and frequency. The arrow to the right of the plot show the range of frequencies observable by a given telescope.}
    \label{fig:Detectability}
\end{figure*}

\section{Results and Discussion}\label{Disc}

\subsection{Comparison to Light Curve Observations, Observed SN Rates, and Progenitor Implications}

The diagnostic values we consider to compare the interaction powered bolometric light curves from our CSM models are the peak luminosity of the light curve and its rise time. The heat map of these values are shown in Figure \ref{fig:Lum_rise}. The peak luminosity and rise times of observed Ibn/Icn SNe lie in the region between 1 day and 20 days, and $10^{42}\,\text{erg}\,\text{s}^{-1}$ and $10^{45}\,\text{erg}\,\text{s}^{-1}$. We consider our models that have peak parameters in this region to be those that would result in a SEISNe. When looking at the peak light curve parameters from our models and comparing them to those of observed SEISNe (Figure \ref{fig:Lum_rise}) the \emph{pure wind} scenario is unable to produce sufficient luminosities at the correct times, and therefore no single star model is able to. This is due to the CSM  density being too low in the \emph{pure wind} scenario. However, both the \emph{wind capture} and \emph{free streaming} scenarios are able to produce peak parameters comparable to those observed. Therefore the models indicate that the wind distributed CSM alone cannot explain the light curves of SEISNe. Previous work by \cite{2022Maeda} has also shown that a $\rho \propto R^{-2}$ distribution, as our wind distribution is in the majority of the models shown in Figure \ref{fig:Density_dis}, is insufficient and that a  steeper density gradient is needed. In the modelling done here the CBD is required to produce the steeper density gradient necessary to explain observed properties of SEISNe, thereby indicating exclusively binary progenitors. We note that other mass-loss mechanisms could also produce a steep density profile (see subsection \ref{Caveats}).\\

In comparison to the \emph{wind capture} scenario, in the \emph{free streaming} scenario the overall expected rate of models that match SEISNe light curve properties are significantly lower, and the range of possible metallicity ZAMS mass combinations is narrower, as shown in Figure \ref{fig:Mass_met}. The expected percentage rates of SEISNe compared to all CCSNe from each scenario are: the \emph{wind capture} scenario - 0.10 per cent, the \emph{free streaming} scenario - 0.026 per cent. The rate from the \emph{wind capture} scenario is consistent with the observed rate of SNe Ibn from \cite{2025Pessi} of $0.45^{+0.44}_{-0.42}$ per cent, and the rate from the \emph{free streaming} scenario is formally consistent with it. \cite{2025Ma} finds the rate of SNe Ibn compared to CCSNe to be $2.1^{+2.0}_{-1.6}$ per cent which is consistent with \cite{2025Pessi} within the lower bounds of their errors. This suggests that a CBD can be the mechanism by which sufficiently dense CSM is present to power SEISNe light curves, whether that CBD captures the wind or not. \\

Our results suggest that the SEISNe progenitors can arise from primary stars in binary systems at $\gtrsim Z_\odot$ with ZAMS masses in the range $\sim14\,\text{M}_{\odot}$ to $\sim40\,\text{M}_{\odot}$ (Figure \ref{fig:Mass_met}). However, at sub-Solar metallicities the ZAMS mass of the primary star needs to be $\sim30\,\text{M}_{\odot}$ to $\sim40\,\text{M}_{\odot}$. The primary star is required to be this massive at these metallicities so it still has sufficient wind driven mass loss to build up the density of the CBD. At $\gtrsim Z_\odot$ the higher metallicity allows for sufficient mass loss due to wind even with a moderately massive mass star.\\

The results of our modelling supports SEISNe being found in star-forming, or high metallicity environments. The characterisation study of \citet{2015Pastorello} showed that the majority of Ibn are found in star-forming regions. However, since then, some Ibn have been found in regions with very low levels of star formation, and/or low metallicity (PS1-12sk: \citealt{2013Sanders}, SN2023tsz: \citealt{2025Warwick}). These regions are unlikely to produce as many $>30\,\text{M}_\odot$ stars as regions with higher star formation rates, and so our modelling predicts SEISNe should be less common within them. Another prediction from our modelling is that the ages of the environments of SEISNe should correlate with the metallicity of the environment. For a sub-solar metallicity environment it should have to be \emph{very} young in order to have $>30\,\text{M}_\odot$ stars exploding within it, due to the lifetimes of these stars. However for a super-solar metallicity environment it can be a bit older as in these environments less massive stars can lead to SEISNe, and these stars will have a longer lifespan than the $>30\,\text{M}_\odot$ stars. \\

Our modelling does suggest that a small number of events, compared to the total SEISNe population, would come from low mass, low metallicity progenitors in both the \emph{wind capture} and \emph{free streaming} scenarios. The $9.0$ and $9.5\,$M$_{\odot}$ Z=$10^{-4}$ region in Figure \ref{fig:Mass_met} is populated by two models, one for each mass. For the $9.5\,$M$_{\odot}$ model it is able to replicate SEISNe light curve features due to it starting with the lowest binary separation of the $9.5\,$M$_{\odot}$ models that pass our initial selection criteria, thereby allowing greater stripping effect from binary interaction. For the $9.0\,$M$_{\odot}$ the sufficient level of binary interaction is due to it having q=0.8. Other models with the same initial parameters as the $9.0\,$M$_{\odot}$ but with higher metallicities also pass our initial selection criteria. However, at these higher metallicities the model results in a merger, meaning there is no late time binary interaction.\\

Currently available optical polarimetry that constraints the geometry of the SN photosphere can be used to evaluate if clearly aspherical CSM  is surrounding the SNe. There are only four SNe Ibn/Icn with conclusive polarimetry, used to probe their geometries: The SNe Ibn 2023tsz \citep{2025Warwick} and 2023emq \citep{2023Pursiainen}, and SN Icn 2021csp \citep{2022Perley}, and the SN Ibn/Icn 2024abvb \citep{2026Anderson} all exhibited low polarisation, while SN Ibn 2015G showed polarisation up to $\approx2.7$ per cent \citep{2017Shivvers}. It should be noted the exact value is unclear due to an uncertain, but possibly substantial, interstellar polarisation (ISP) contribution. While the sample is still small, it suggests most SEISNe appear to be consistent with a high degree of spherical symmetry, in opposition to what our models suggest. It should be noted that in a CBD model the polarisation could still appear low depending on the viewing angle of the system. The SN blast wave will have passed the disk in $<8$\,days in the majority of our models. This means that high polarisation from when the photosphere is formed in the nearby disk CSM could give way to low polarisation once the photosphere moves beyond the disk. Polarisation consistent with this was seen in 2024abvb which showed polarisation of $\sim1$ per cent near peak and then $\lesssim0.5$ per cent a week later \citep{2026Anderson}. The rest of the currently available polarimetry of Ibn/Icn SNe does likely not capture the early phases when the photosphere is formed in the nearby disk CSM, and high early polarisation may have been missed. As such the polarimetry is consistent with the derived CSM properties here.\\

\subsection{Comparison to Radio Observations}

\subsubsection{Observations of CSM density}

Figure \ref{fig:Density_dis} shows the CSM from our modelling compared to radio observations probing the CSM from SE SNe. When comparing our CSM models to radio observations we only consider the CSM beyond $10^{15}$\,cm. This is because radio observations have only directly measured the CSM of SEISNe beyond this radius as observations within a few days of explosion are required to probe smaller radii. Additionally we only consider the CSM density up to $3 \times 10^{17}\,$cm as this is the extent of the observations from \cite{2025Sfaradi}. We note that there are radio observations of SE SNe taken at later times that probe larger radial distances from the SNe \citep[e.g. ][]{2021Stroh, 2024Rose}. However, these were typically single  band observations that do not provide strong constraints on the environment densities. At radii greater than $10^{15}\,$cm the density is primarily comprised of mass lost through winds. This density behaviour is consistent across model types -1, 0, and 1 (singles, mergers, and binaries). In Figure \ref{fig:Density_dis} there are isolated pixels, these are caused by the discrete time steps of the BPASS models leading to non-continuous CSMs that can have large density differences between steps. Additionally, when a higher velocity shell of CSM has caught up to a lower velocity shell of an earlier time step an increase in density is seen at the radius where the shells overlap, noticeably higher the the preceding and following shells.\\

The CSM densities from all of the models we have selected from BPASS agree closely with the observed CSM densities from \cite{2025Sfaradi}. The radius range probed by the observations are all above the median outer edge of our expected CBD with the majority being further out than the maximum disk sizes from our models. Current radio observations are therefore likely being taken too late to see the interaction with any CBD.  Our modelling suggests that at the time of the observations, the CSM the blast wave is interacting with is dominated by that produced by the wind driven mass loss of the progenitors. Therefore they provide no constraints on any RLOF driven mass loss. Our models also show that, whilst less likely, the mass loss from RLOF is sufficient to explain observed CSM densities if distributed by the winds.\\

A notable output of our models, in terms of CSM distribution, is a region of low occurrence rate but high density CSM distinct from the majority of models, $\frac{\rho}{\mathrm{g\,cm^{-3}}} \gtrsim 10^{20} \left(\frac{R}{\mathrm{cm}}\right)^{-2.4}$ (see Figure \ref{fig:Density_dis}). The models that populate this region are all single star models formed from mergers at higher metallicities. This aligns with what we would expect as mergers are likely to produce higher mass WR stars that have denser winds, exploding as WNs rather than WCs. \\

\subsubsection{Detectability of Synchrotron Emission from the Median CBD}

When modelling the synchrotron emission from a CBD we use the median disk parameters from our models. These median disk parameters are a density of $1.2 \times 10^{-13}\,\text{g}\,\text{cm}^{3}$, an inner disk radius of $8.4 \times 10^{12}\,\text{cm}$, and an outer disk radius of $2.8 \times 10^{14}\,\text{cm}$. A shock wave with a constant velocity of $4345\,\text{km}\,\text{s}^{-1}$ (the median ejecta velocity from our models), would pass through a disk of this size in $\approx 7.6\,$days. Our upper disk size is $4.0\times10^{15}\,\text{cm}$, and for this upper limit the shock wave would pass through in $\approx 105.9\,$days. The observability of the synchrotron emission from a disk with these parameters, taking into account synchrotron self absorption but not FFA, is shown in Figure \ref{fig:Detectability}. Taking into account FFA we find the disk to be opaque to the radio emission, and the wind transparent (at $\geq3\,\text{GHz}$). Therefore it is expected emission would be visible from interaction of the shock front with the disk only for viewing angles out of the plane of the disk. At SN distances of up to $100$\,Mpc, ALMA observations at $\lesssim8$\,d would be able to detect the interaction signature of the blast wave as it moves through disks above the 70th percentile in size. At 500\,Mpc only the largest disks would be detectable during this phase.\\

From our modelling of the synchrotron emission from the median CBD we find that high frequency radio observations taken within a few days of outburst would be needed to observe the majority of the disks. The SN would also need to be within 100\,Mpc for the radio flux density to be above the detection limits, and the viewing angle would need to be away from the plane of the disk to avoid FFA by the disk. Radio observations of close systems ($<50\,$Mpc) taken within 8 days would be able to detect the majority of disks, and therefore be able to confirm the presence, and density of a CBD. A non-detection in this time would not necessarily rule out a CBD, as the viewing angle could mean the disk could be obscuring the radio emission. Concurrent polarimetry would therefore be useful, as in this situation we would expect a high degree of asymmetry. \\

The radio emission from interacting supernovae has been investigated before by \cite{2025Wu}. Their modelled CSM densities are higher than those found in our models. However, this difference is expected given they predict detached disks in their modelling that our models do not. Additionally, their work is focused on late-time radio emission as they predict the detached disks at large radii. We focus primarily on earlier radio emission from a compact CBD. A direct comparison of radio spectra is also not possible, as they do not present full radio SEDs. As such, our work should be viewed as probing complementary regimes of CSM interaction to their analysis rather than being a direct comparison.

\section{Modelling Caveats} \label{Caveats}

In the work described in this paper we have combined BPASS stellar evolution models with simplified prescriptions for the distribution of CSM and the conversion of energy from the ejecta-CSM interaction into observable emission, in order to assess whether these systems can reproduce the key properties of observed SEISNe. This approach necessarily relies on a number of assumptions and idealisations, both in how the CSM is constructed and in how the resulting interaction is modelled. In this section we therefore discuss the main caveats of our modelling, focusing on the assumptions that are most relevant to the inferred luminosities, timescales, and event rates. We also highlight several physical processes that are not explicitly included in our framework, and consider how their inclusion might alter or qualify the conclusions drawn from our results.\\

In our models we have assumed the conversion of the kinetic energy to observed luminosity in the blast wave is 100 per cent. Figure 11 in \cite{2024Khatami} shows that for the CSM masses and radii relevant to our models the efficiency would be in the range 10 to 50 per cent. We are optimistic with the efficiency value we use, but conservative in other areas. Firstly we are looking at CBD and CSM shells of constant density, whereas these will likely have regions of higher and lower density within them. Particularly in the case of the CBD, localised regions of higher density may enhance the peak luminosity, especially in more extreme configurations. Additionally, increases in the opening angle of the CBD above $10^\circ$ would result in a greater fraction, and so mass, of the ejecta interacting with it, further increasing the peak luminosity. For our models producing SEISNe parameters the average mass ratio of the binary is 0.5, \cite{2025Scherbak} suggests an opening angle of $15^\circ$ for this mass ratio, with some of our models having mass ratios suggesting upwards of $30^\circ$ for their opening angles. Our 100 per cent efficiency and $10^\circ$ together compensate for a smaller conversion efficiency and higher opening angle.\\ 


We have assumed in our CBD model that the disk has constant density. However, \cite{2023Tuna} show that the surface density decreases as a function of distance. Therefore, the density of the disk will be greatest at its inner edge, and this density would be greater than the disk average. This would have the effect of increasing the peak luminosities and reducing the peak times of the SNe light curves, shifting the parameter region in the lower two panels of Figure \ref{fig:Lum_rise} up and to the left. Whilst this would change the models that lie in the SEISNe peak parameter region it would have a minimal effect on the event rate. This is due to the section of the heatmap that would now occupy this region having a comparable density to the section currently occupying the region.\\

Our $^{56}$Ni mass to ejecta mass of 1:14 is likely too high for SEISNe. There is currently uncertainty on the likely ratio due to the difficulty of observing the later stages of the SN light curve and constraining the tail of the $^{56}$Ni decay (As was the case in \citealt{2023Pursiainen} and \citealt{2025Warwick}). However, previous work has shown very low $^{56}$Ni masses \citep{2016Moriya, 2022Perly}. Given the peak parameters of our models are at the interaction peak, and that at these times the luminosity is already dominated by the interaction power in SEISNe, it is unlikely that the reduction in $^{56}$Ni mass would have a notable impact on our peak luminosity. \\

In our modelling upon explosion we do not continue to take into account the expansion of the wind-driven CSM, although we do still consider the velocity difference between the CSM and the blast wave. The effect of the CSM continuing to expand after explosion would be to delay the peak of the interaction in the \emph{pure wind} scenario. In both of the CBD scenarios there would be minimal effect on the peak time as the velocity of the blast wave is significantly larger than that of the wind. In Figure \ref{fig:Lum_rise} the plot for the \emph{pure wind} senario does show a region of models producing SEISNe luminosities, but occurring sooner than SEISNe timescales. In order for this region to occur at SEISNe peak times an expanding wind would have to be moving over 99 per cent the velocity of the ejecta. Therefore having the wind be static after explosion doesn't meaningfully affect our results.\\

For all of our models we treat their wind speeds as those of a WR star. However, a number of our models will be helium giants (those with a final mass of $\lesssim 6\,\text{M}_{\odot}$). These stars will have a lower wind speed. Their mass-loss rates are also expected to differ from those adopted in the current BPASS version, and are likely to be lower \citep{2017Vink}. The combined affect of these two factors would mean a lower mass CSM closer to the star. These could work in opposition in the \emph{pure wind} scenario for both peak luminosity and time, as despite this CSM having lower mass it would denser as it wouldn't have spread out as far. For the \emph{free streaming} scenario, as the disk is formed of mass lost due to RLOF again we would not expect a notable change to the peak luminosity or time. This means the \emph{free streaming} scenario sets a lower bound for our expected SEISNe rate from the \emph{wind capture} scenario as in a zero wind scenario they would be identical. However, for the \emph{wind capture} scenario it would change the amount of material available to be captured by the disk, possibly lowering its peak luminosity. In terms of rates, models ending in stars of mass $\lesssim 6\,\text{M}_{\odot}$ make up 84 per cent of the total expected SEISNe. So correct helium giant mass loss rates would likely reduce our expected SEISNe from our models, and change the metallicity and initial mass model parameters to those seen in the \emph{free streaming} scenario. \\

One consideration that isn't included in our models are periods of eruptive mass loss in the progenitors. Such periods could result in higher mass loss rates at the end of stars life, leading to a close-in dense CSM without the need for a CBD. This type of mass loss has been observed in several SNe \citep[e.g.][]{2024Elias-Rosa, 2025Gkini, 2025Nagao}. The stars that undergo this late stage mass loss are though to be massive stars ($>20\,\text{M}_\odot$) such as luminous blue variables \citep{2013Groh} but could also occur from lower mass progenitors that undergo core collapse, those that end their lives with CO core masses of $1.38 - 1.68\,\text{M}_{\odot}$ \citep{2015Woosley}. We therefore note that eruptive mass loss provides a plausible alternative pathway for producing a dense, close-in CSM and should be considered when interpreting our results. \\

\section{Conclusions}\label{Con}
Using BPASS stellar evolution models we have investigated the circumstellar material (CSM) and observable resulting SN properties of stripped envelop interacting supernovae (SEISNe) progenitors. Our key findings are summarised as follows:

\begin{enumerate}
    \item \textbf{Binary interaction can reproduce the observed features of SEISNe.} Pure wind models cannot generate sufficiently dense close-in CSM to match the observed peak light curve properties of Ibn/Icn. CSM confined into a circum-binary disc (CBD) by binary interactions provides a mechanism to have dense enough material close to the progenitor. However, we can not definitively state that binary interactions are required for SEISNe as other phenomena, such as eruptive mass loss, may also be able to reproduce the observed features.
    \item \textbf{To detect CBDs we require high-frequency radio observations within days of the supernova.} Given the expected size and density of the CBDs, detections of the synchrotron emission from the interaction region between it and the SN ejecta would require observations taken in the first 8 days. Even so, free-free absorption (FFA) by the CBD would mean that the disc signature would only detectable at viewing angles outside the plane of the disc.
    \item \textbf{Expected progenitors of SEISNe are massive stars in binary systems.} The majority of SEISNe progenitors are expected to be stars in binary systems with ZAMS masses of $14-40\,\text{M}_{\odot}$ at and above solar metallicities. Some are expected to come from sub-solar metallicity progenitors with ZAMS masses of $30-40\,\text{M}_{\odot}$. However, at sub solar metallicities very few models with $<20\,\text{M}_{\odot}$ are able to reproduce the observed peak light curve properties of SEISNe. This means that the majority of SEISNe are expected in high star formation rate and high metallicity host environments. 
\end{enumerate}

Overall, our results indicate that stripped-envelope interacting SNe are best explained by massive binary progenitors undergoing mass transfer in the final $100\,$kyrs, where a CBD captures and concentrates material close to the progenitor prior to core collapse. This configuration reproduces both the luminosity and timescale of observed Ibn/Icn events, and predicts a rate consistent with the known population. Early-time radio follow-up and polarimetric constraints will be crucial in confirming the presence of the CBD, and its properties. This predicts that SEISNe should preferentially occur in environments that efficiently produce massive binaries, favouring regions of high star formation and near-solar or super-solar metallicity, with lower-metallicity events arising only from the most massive progenitors.

\section*{Acknowledgements}

BW acknowledges the UKRI's STFC studentship grant funding, project reference ST/X508871/1. JDL and MP acknowledge support from a UK Research and Innovation Future Leaders Fellowship (grant references MR/T020784/1 and UKRI1062). DLC, CMB and ERS acknowledge support from the Science and Technology Facilities Council (STFC) grant number ST/X001121/1. Computing facilities were provided by the Scientific Computing Research Technology Platform of the University of Warwick. We would like to thank the anonymous reviewer for their helpful comments and feedback.

\section*{Data Availability}

The underlying BPASS models are available at \url{https://warwick.ac.uk/fac/sci/physics/research/astro/research/catalogues/bpass/}. CSM models and data included in figures and tables is available upon reasonable request to the author.



\bibliographystyle{mnras}
\bibliography{biblo} 




\appendix 

\section{Binary Interaction Cases} \label{Cas}
We invesigate our selected BPASS models to see which undergo mass transfer due to RLOF or common envelope evolution (CEE). We also investigate the nature of this mass transfer and consider four different cases:\\
\begin{itemize}
    \item \textbf{Case A}: Mass transfer begins while the donor star is still in its core hydrogen-burning (main sequence) phase.
    \item \textbf{Case B}: Mass transfer starts after the donor leaves the main sequence but before helium ignition (during shell hydrogen burning).
    \item \textbf{Case BB}: A second phase of mass transfer from a stripped helium star after earlier mass loss (post–Case B evolution).
    \item \textbf{Case C}: Mass transfer occurs after the donor has ignited helium, typically during its late giant (asymptotic giant branch) phase.
\end{itemize}

\begin{figure*}
    \centering \includegraphics[width=\textwidth,height=0.9\textheight,keepaspectratio]{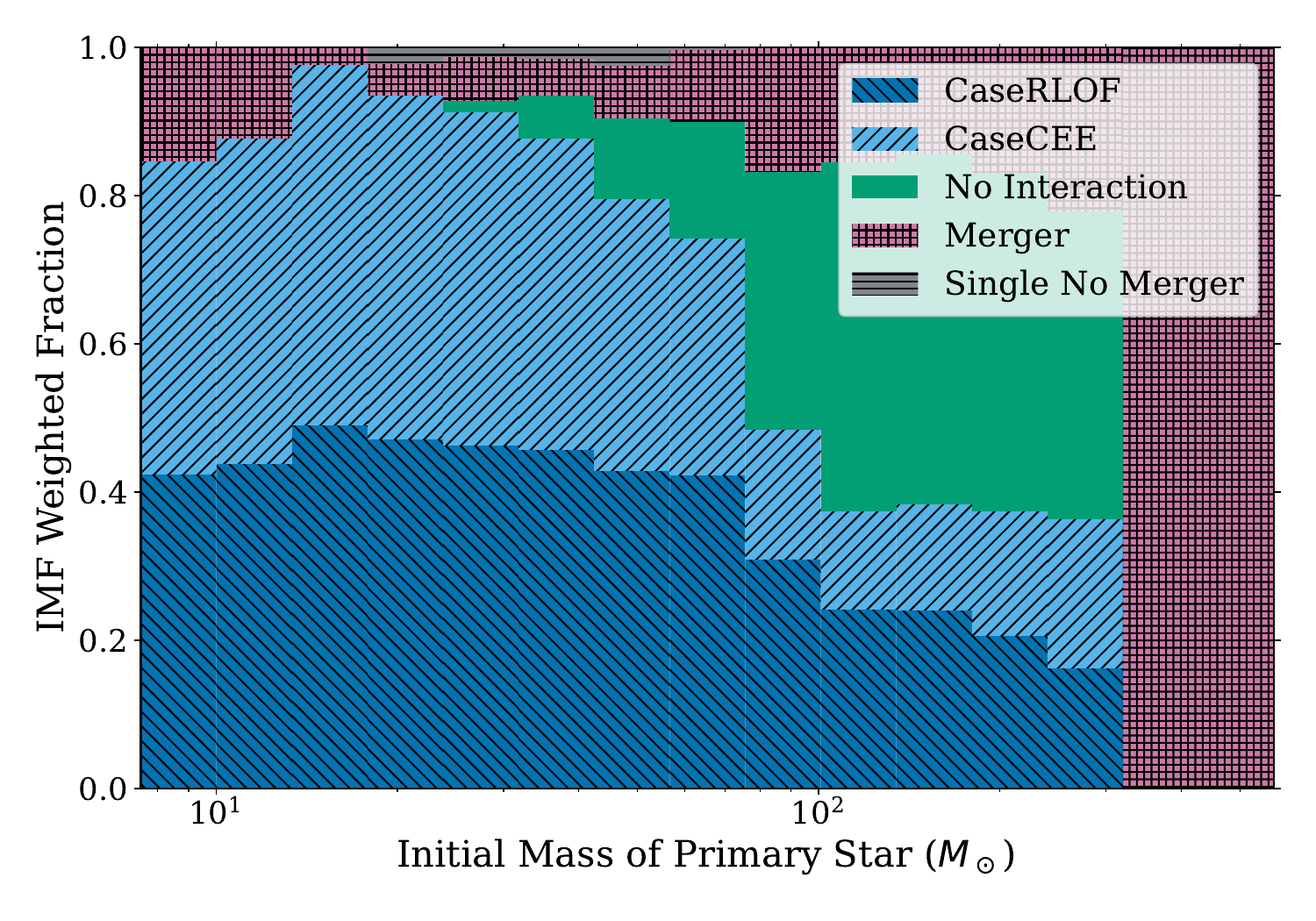}
    \caption{The IMF-weighted fraction of the selected BPASS models that undergo binary interactions, at some point over the entire lifetime of the primary star, as a function of the initial stellar mass of the primary star. The 'No Interaction' solid colour region shows the parameter space occupied by binary models that have no binary interaction, with the 'CaseRLOF' (backslash hatched) and 'CaseCEE' (Common Envelope Evolution) (forward slash hatched) regions showing the parameter space of those models that undergo Roche lobe overflow and common envelope evolution respectively. The plus hatched 'Merger' region  shows the parameter space occupied by models that are single stars that have resulted from a merger, whilst the horizontal line hatched 'Single No Merger' region shows the parameter space for the remaining non-merger single star models.}
    \label{fig:Bin_type}
\end{figure*}

\begin{figure*}
    \centering \includegraphics[width=\textwidth,height=0.9\textheight,keepaspectratio]{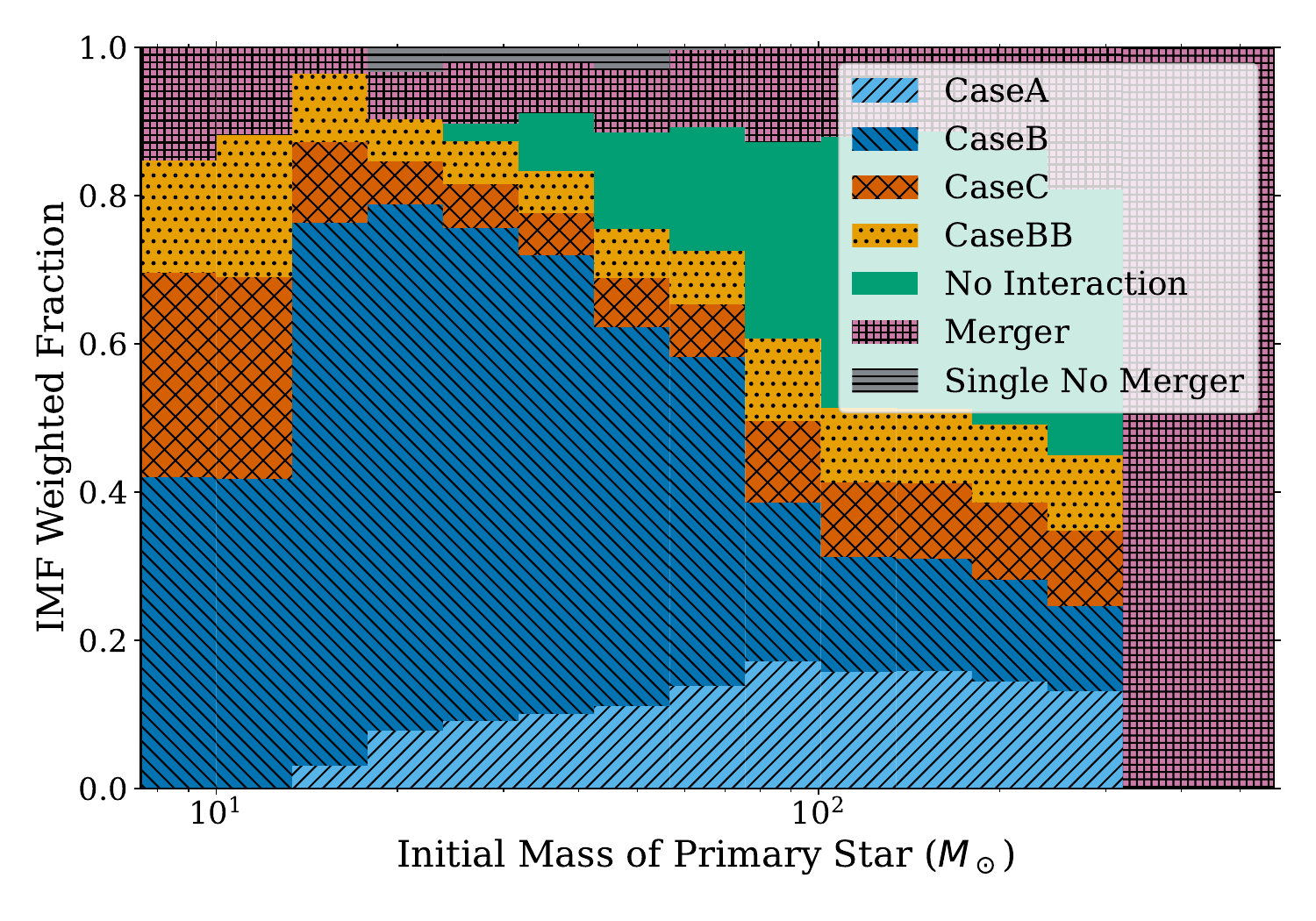}
    \caption{The IMF-weighted fraction of the selected BPASS models that undergo case A, B, BB, and C mass transfer at some point over the entire lifetime of the primary star, as a function of the initial stellar mass of the primary star. The different models refer to the evolutionary stage where mass transfer commenced; namely on the main-sequence (Case A), after the main-sequence but prior to He ignition (case B), a second phase of mass transfer prior to He ignition (Case C), or mass transfer after He ignition (Case C). See the text in \ref{Cas} for more details. The 'No Interaction' solid colour region shows the parameter space occupied by binary models that have no binary interaction, with the 'Case A' (forward slash hatched), 'Case B' (backslash hatched), 'Case C' (cross hatched), and 'Case BB' (dot hatched) regions showing the parameter space of those models that undergo Case A, Case B, Case BB, and Case C respectively. The plus hatched 'Merger' region  shows the parameter space occupied by models that are single stars that have resulted from a merger, whilst the horizontal line hatched 'Single No Merger' region shows the parameter space for the remaining non-merger single star models.}
    \label{fig:Case_type}
\end{figure*}

\begin{figure*}
    \centering \includegraphics[width=\textwidth,height=0.9\textheight,keepaspectratio]{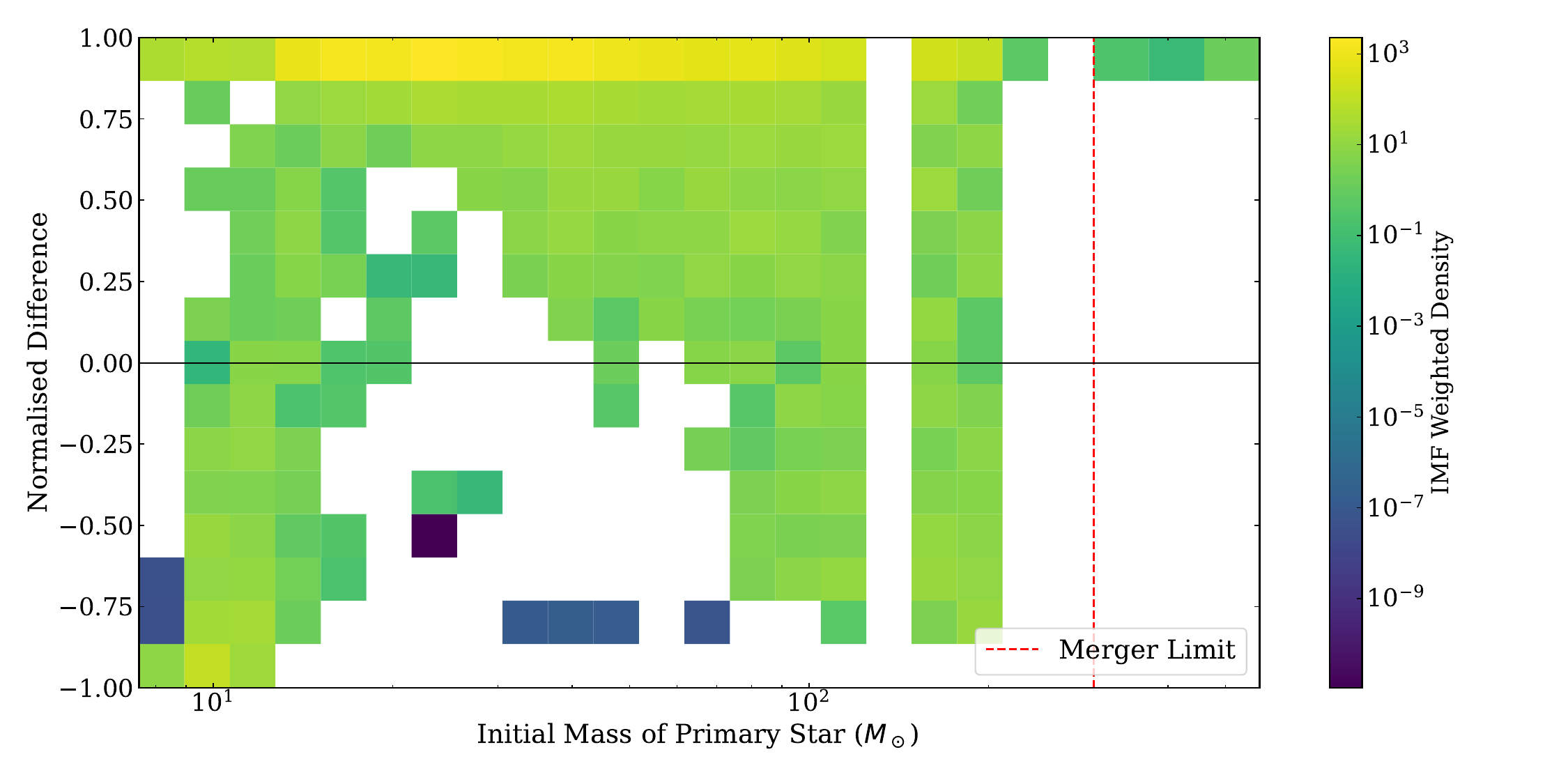}
    \caption{The IMF-weighted density of the mass lost through wind compared to mass lost through RLOF, for our selected BPASS models, in the last 100\,kyr of their evolution by ZAMS of the primary star. The comparison is shown as the normalised difference between the wind and RLOF contribution where 1 is pure wind mass loss, and -1 is pure RLOF mass loss. The y-axis is calculated as (a-b) / (a+b), where a is the mass lost through wind and b is the mass lost through RLOF. The vertical dashed line shows the limit above which all the models are single star models that formed as a result of a merger.}
    \label{fig:WindRLOF}
\end{figure*}

\begin{figure*}
    \centering \includegraphics[width=\textwidth,height=0.9\textheight,keepaspectratio]{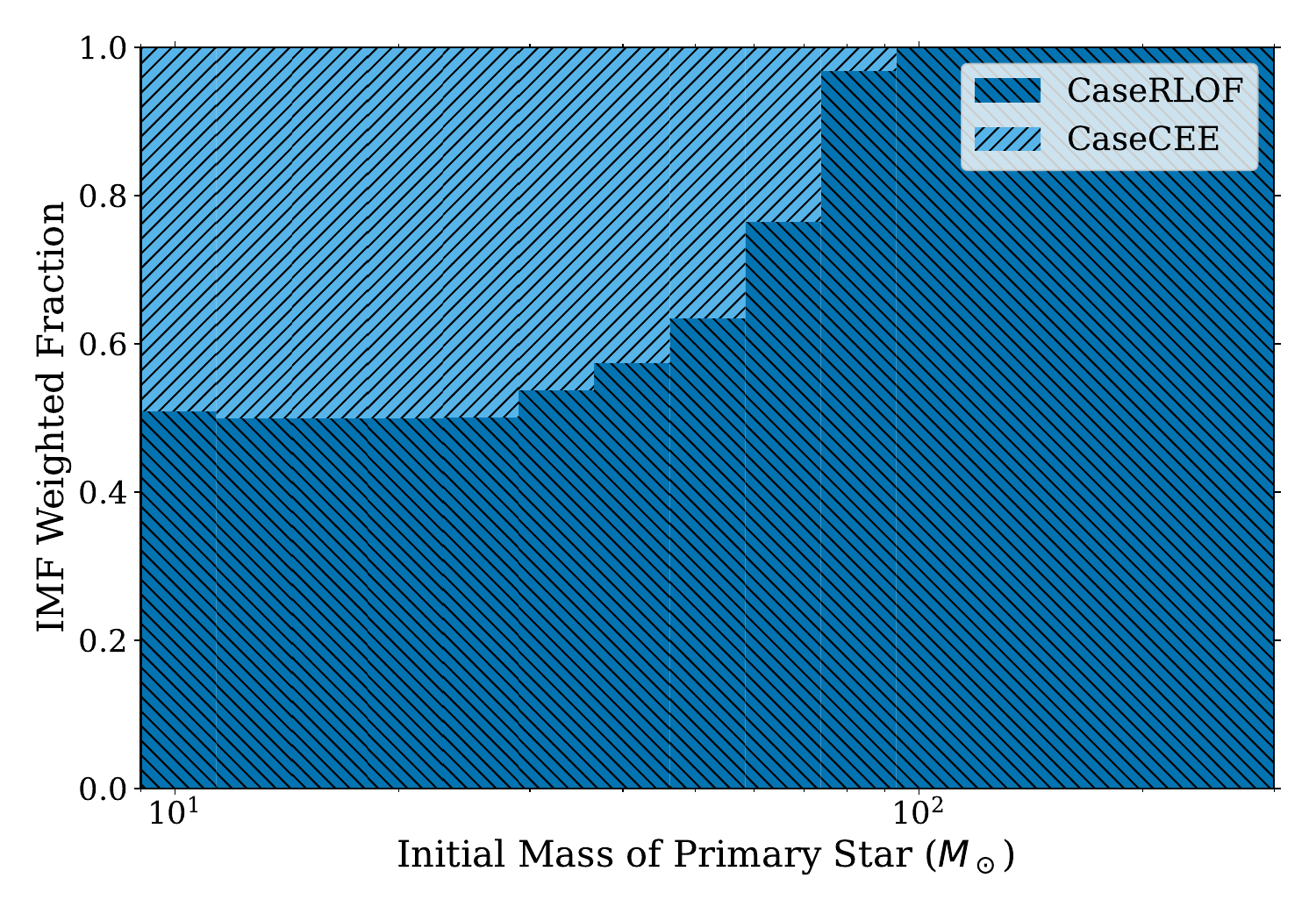}
    \caption{The IMF-weighted fraction of the models we determine to be SEISNe progenitors that undergo binary interactions at some point over the entire lifetime of the primary star, as a function of the initial stellar mass of the primary star. The 'CaseRLOF' (backslash hatched) and 'CaseCEE' (Common Envelope Evolution) (forward slash hatched) regions show the parameter space of those models that undergo Roche lobe overflow and common envelope evolution respectively.}
    \label{fig:Bin_type_SEISNe}
\end{figure*}

\begin{figure*}
    \centering \includegraphics[width=\textwidth,height=0.9\textheight,keepaspectratio]{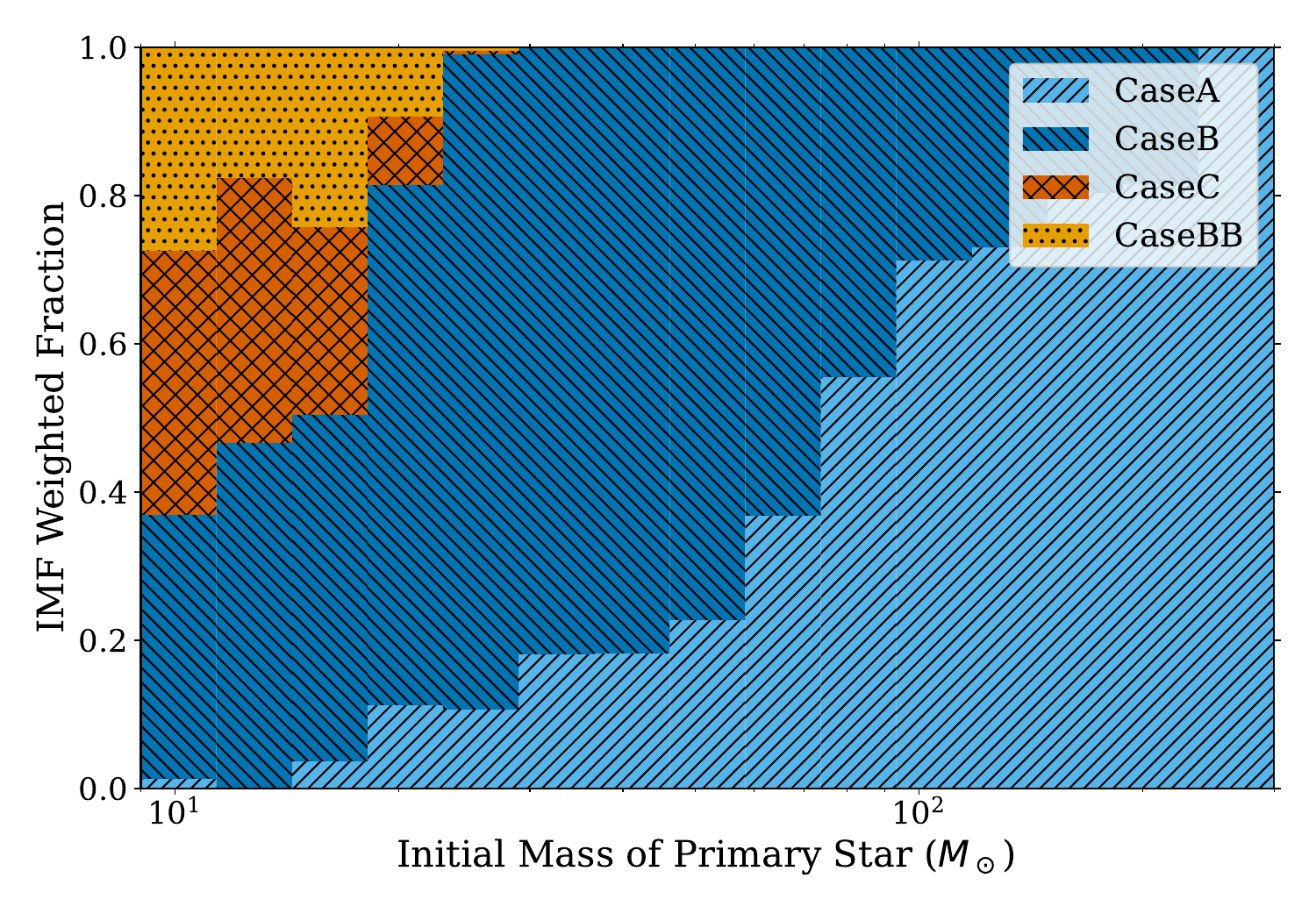}
    \caption{The IMF-weighted fraction of the models we determine to be SEISNe that undergo case A, B, BB, and C mass transfer, at some point over the entire lifetime of the primary star, as a function of the initial stellar mass of the primary star. See the text in Appendix \ref{Cas} for a description of these models.  The 'Case A' (forward slash hatched), 'Case B' (backslash hatched), 'Case C' (cross hatched), and 'Case BB' (dot hatched) regions show the parameter space of those models that undergo Case A, Case B, Case C, and Case BB respectively.}
    \label{fig:Case_type_SEISNe}
\end{figure*}

\begin{figure*}
    \centering \includegraphics[width=\textwidth,height=0.9\textheight,keepaspectratio]{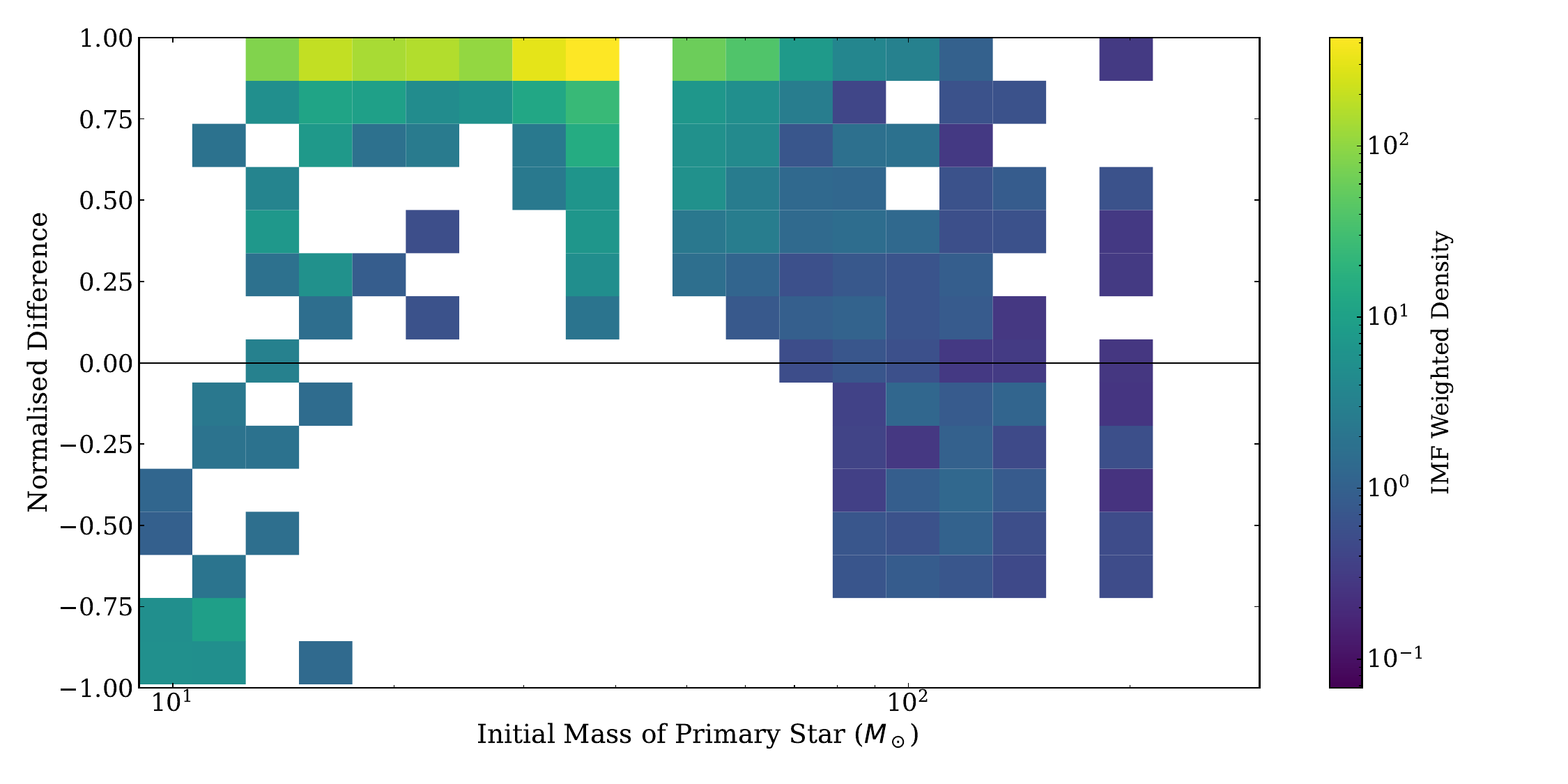}
    \caption{The IMF-weighted density of the mass lost through wind compared to mass lost through RLOF, of the models we determine to be SEISNe, in the last 100\,kyr of their evolution by ZAMS of the primary star. The comparison is shown as the normalised difference between the wind and RLOF contribution where 1 is pure wind mass loss, and -1 is pure RLOF mass loss. The y-axis is calculated as (a-b) / (a+b), where a is the mass lost through wind and b is the mass lost through RLOF.}
    \label{fig:WindRLOFSEISNe}
\end{figure*}

We show the IMF weighted fractions that undergo RLOF and CEE in Figure \ref{fig:Bin_type}, and those that undergo case A, B, BB, and C in Figure \ref{fig:Case_type}. Additionally we show the normalised difference between wind and RLOF mass loss in the last 100\,kyr of the models evolution in Figure \ref{fig:WindRLOF}. Figures \ref{fig:Bin_type_SEISNe}, \ref{fig:Case_type_SEISNe}, and \ref{fig:WindRLOFSEISNe} are comparable plots, but only showing the subset of models that produce SEISNe.\\

\section{Interaction Powered Luminosity} \label{Equ}

We assume the observed luminosity arises from the conversion of the kinetic energy lost by the CDS into
radiation such that,\\

\begin{equation}
    L_{avg, shell} = e\frac{dE}{dt}
\end{equation}\\

where $L_{avg, shell}$ is the average observed luminosity over the time increment $dt$, which is the time the CDS is interacting with a given shell of CSM, with $dE$ being the energy lost by the CDS in the increment, and $e$ being the efficiency of the conversion. We assume the conversion to be 100 per cent efficient, and so $e = 1$. Assuming constant de-acceleration by the CDS as it interacts with a CSM shell of width $dr$ then,\\

\begin{equation}
    dt = \frac{dr}{v_{0, shell}+\frac{dv}{2}}
\end{equation}\\

where $v_{0, shell}$ is the velocity of the CDS when it encounters the shell of CSM, and $dv$ is the change in velocity of the CDS after interacting with the shell of CSM. $v_{0, shell}+\frac{dv}{2}$ therefore represents the average velocity of the CDS as it interacts with a given shell of CSM. We assume the CDS collides inelastically with the shell of CSM, conserving momentum, such that,\\

\begin{equation}
    dv = \frac{M_{0, shell}v_{0, shell}}{M_{0, shell} + dm} - v_{0, shell}
\end{equation}\\

with $dm$ being the mass CSM swept up by the CDS. Under our assumptions the amount of kinetic energy that is converted into observed radiation, $dE$, as the CDS collides with a defined shell of CSM is,\\

\begin{equation}
\begin{split}
    dE = \frac{1}{2}M_{0, shell}v_{0, shell}^2 - \frac{1}{2}(M_{0, shell} + dm)(v_{0, shell} + dv)^2 \\
    = -\frac{1}{2}M_{0, shell}v_{0, shell}dv
\end{split}
\end{equation}\\

the observed luminosity is therefore,\\

\begin{equation}
\begin{split}
    L_{avg, shell} = -\frac{1}{2}M_{0, shell}v_{0, shell}dv\frac{v_{0, shell} + \frac{dv}{2}}{dr} \\
    = \frac{1}{4dr}M_{0, shell}v_{0, shell}^3\left[1-\frac{M_{0, shell}^2}{(M_{0, shell}+dm)^2}\right]
\end{split}
\end{equation}\\


\bsp	
\label{lastpage}
\end{document}